\documentclass[aps,pre,twocolumn,floatfix,superscriptaddress,showpacs]{revtex4-1}

\usepackage{amsmath}
\usepackage{amssymb}
\usepackage{bbold}
\usepackage{graphicx}
\usepackage{hyperref}

		\def \pvec {\mathbf{p}}
		\def \cmax {c_\textrm{max}}
		\def \cvmin {\textrm{CV}_\textrm{min}}

\makeatletter
\renewcommand*\env@matrix[1][\arraystretch]{%
  \edef\arraystretch{#1}%
  \hskip -\arraycolsep
  \let\@ifnextchar\new@ifnextchar
  \array{*\c@MaxMatrixCols c}}
\makeatother

\begin{document}


\title{Stochastic modeling of cell growth with symmetric or asymmetric division}

\author{Andrew Marantan}
\affiliation{Department of Physics, Harvard University, Cambridge, MA 02138}
\author{Ariel Amir}
\affiliation{School of Engineering and Applied Sciences, Harvard University, Cambridge, MA 02138}

\begin{abstract}
	We consider a class of biologically-motivated stochastic processes in which a unicellular organism divides its resources (volume or damaged proteins, in particular) symmetrically or asymmetrically between its progeny. Assuming the final amount of the resource is controlled by a growth policy and subject to additive and multiplicative noise, we derive the ``master equation'' describing how the resource distribution evolves over subsequent generations and use it to study the properties of stable resource distributions. We find conditions under which a unique stable resource distribution exists and calculate its moments for the class of \emph{affine linear} growth policies. Moreover, we apply an asymptotic analysis to elucidate the conditions under which the stable distribution (when it exists) has a power-law tail. Finally, we use the results of this asymptotic analysis along with the moment equations to draw a stability phase diagram for the system that reveals the counterintuitive result that asymmetry serves to increase stability while at the same time widening the stable distribution. We also briefly discuss how cells can divide damaged proteins asymmetrically between their progeny as a form of damage control. In the appendix, motivated by the asymmetric division of cell volume in \emph{Saccharomyces cerevisiae}, we extend our results to the case wherein mother and daughter cells follow different growth policies.
\end{abstract}


\maketitle

\section{Introduction}

	When a unicellular organism divides, it allocates its cellular resources (proteins, DNA, etc.) to its newborn daughter cells. Though the parent cell often distributes many of these resources equally (as occurs in prokaryotes like \emph{Escherichia coli} or eukaryotes like \emph{Schizosaccharomyces pombe}), there are exceptions. For example, the yeast \emph{Saccharomyces cerevisiae} divides by budding, which results in the budded daughter cell inheriting a smaller volume. In other cases this asymmetric allocation serves a more obvious purpose, as when a parent cell actively segregates its damaged proteins into one of its (ill-fated) daughter cells \cite{sinclair1997extrachromosomal, aguilaniu2003asymmetric, winkler2010quantitative} to ensure the other survives or when a mother cell keeps most of its stores of a scarce resource so that it may continue to proliferate \cite{avraham2013increasing}. Whether it be volume, proteins or another resource, the population-level distribution for a given resource tends to stabilize over successive generations and the amount of symmetry or asymmetry in the division of that resource nontrivially affects the stable distribution's statistical properties.

	While in previous studies we focused (theoretically and experimentally) on the correlations between various cell cycle variables \cite{amir2014cell, brenner2015universal, ho2015simultaneous}, we devote this study to a quantitative, statistical analysis of stable resource distributions. Assuming that every cell in the population attempts to accrue resources according to a species-specific growth policy that takes into account the initial amount of the resource at birth, $v$, and specifies the desired (pre-division) amount $v^G = f(v)$, we find the conditions under which a unique stable resource distribution exists and study its properties. In addition, we allow for multiplicative and additive noise during growth (see Sec. \ref{Sec:Model_Spec}) and assume that the asymmetry ratio $r$ (the ratio of the resources allocated to the two offspring) is fixed \cite{marr1969growth, mannik2012robustness, robert2014division, soifer2016single} and leave the more general case to App. \ref{App:Stochastic_Division}.

	More formally, we show that the distribution for the amount of resources a cell born into the $n$th generation has at birth, $P_n(v)$, evolves into the distribution for the next generation $P_{n+1}(v)$ according to an (integral) master equation (see Sec. \ref{Sec:Model_Building}),
	
	\begin{equation}
		P_{n+1}\left( v \right) = \int_0^\infty dv' \; K(v, v') \, P_n\left( v' \right),
	\label{eq:Recursive_Relation}
	\end{equation}
	
	\noindent in which the growth policy $f(v)$ determines the kernel of the integral, $K(v, v')$. Then by taking $P_{n+1}(v) = P_n(v) = P(v)$, we obtain a homogenous Fredholm integral equation of the second kind,
	
	\begin{equation}
		P(v) = \int_0^\infty dv' \; K(v, v') \, P( v' ),
	\label{eq:Eigenvalue_Equation}
	\end{equation}
	
	\noindent which provides a necessary condition for a stable resource distribution that allows us to address questions of existence and uniqueness (Sec. \ref{Sec:Proof}). Note that our choice to work with cell generations instead of time is for analytical convenience: the two approaches are equivalent with regards to understanding stability (see App. \ref{App:Generations_vs_Time}). 
	
	Given the abundance of recent studies experimentally probing cell size distributions by tracking cell volume at the single-cell level, both for symmetrically dividing cells \cite{amir2014cell, campos2014constant, taheri2015cell, deforet2015cell, kennard2016individuality} and asymmetrically dividing cells \cite{soifer2016single, iyer2014scaling}, we couch the majority of our discussion of the stable resource distribution in terms of the cell volume distribution in symmetrically- or asymmetrically-dividing cells. Other recent studies looked at protein number distributions at the single-cell level \cite{brenner2006dynamics, salman2012universal, brenner2015universal}, and so we also pay special attention to asymmetric division of damaged proteins in Sec. \ref{Sec:Damage_Control}. Nevertheless, the mathematical results we derive apply equally well to other resource distribution problems.
	
	In addition to the evolution and consistency equations (Eqs. \ref{eq:Recursive_Relation} and \ref{eq:Eigenvalue_Equation}), we present here three main results for the case of affine linear growth policies:
	
	\begin{enumerate}
	
		\item explicit formulae for the stable volume and damage distributions' mean and variance in terms of the policy and noise parameters (Secs. \ref{Sec:Moments} and \ref{Sec:Damage_Control}),
		\item an asymptotic analysis that reveals the stable distribution's power-law tail and provides an equation for the tail's power (Sec. \ref{Sec:Tail}),
		\item a stability phase diagram characterizing the range of parameters for which a stable distribution exists (Sec. \ref{Sec:Phase_Diagram}) that suggests that asymmetric division actually improves stability.
	
	\end{enumerate}
	
	\noindent We also extend our results on the moments of the stable distribution to a model in which mother and daughter cells follow different growth policies (see App. \ref{App:Mother_Daughter_Model}).


\section{Model Specification}
\label{Sec:Model_Spec}

	\begin{figure}[t]
		\centering
			\includegraphics[width=\linewidth]{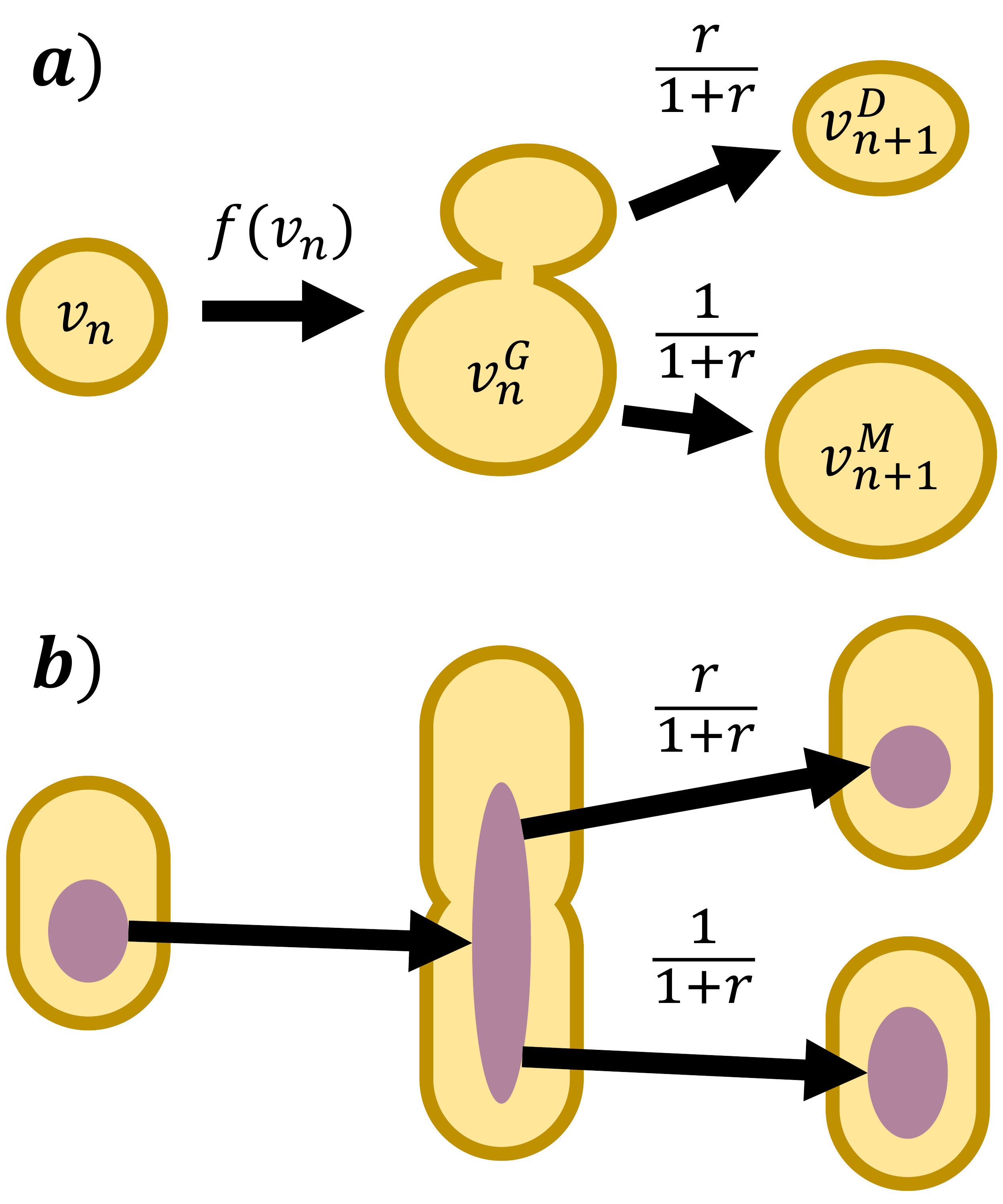}
		\caption{(Color online) \textbf{(a)} Schematic for asymmetric division (budding) in yeast. We consider the growth and division of a cell born with volume $v_n$ at the beginning of the $n$th generation. While this cell grows exponentially in time, $v_n(t) = v_n \, e^{\lambda t}$, a new cell starts to emerge from it as a bud. After growing to a total (i.e. both mother and bud) volume $v_n^G$ according to a growth policy $v_n^G = f(v)$, the budding daughter cell detaches from the larger mother cell. The volume of the daughter cell after division, $v_{n+1}^D$ is related to the volume of the mother cell after division, $v_{n+1}^M$, by a fixed ratio $r$, $v_{n+1}^D = r  \, v_{n+1}^M$. \textbf{(b)} Our model also applies to cases of asymmetric resource division in symmetrically-dividing cells.}
	\label{fig:Asymmetric_Division}
	\end{figure}

	We begin building our stochastic model by discussing how cells of the asymmetrically-dividing yeast \emph{S. cerevisiae} grow and divide. An $n$th generation cell born with volume $v_n$ will grow exponentially in time \cite{elliott1978rate, di2007effects, godin2010using, wang2010robust, turner2012cell} to a fully-grown size $v_n^G$, at which point the daughter cell buds off the mother cell (note that for organisms like \emph{E. coli} which divide symmetrically, we would say the original ``mother'' cell produces two ``daughters''). Since volume is conserved, the birth volumes of the mother and daughter cells, which we denote by $v_{n+1}^M$ and $v_{n+1}^D$, must add up to $v_n^G$, 
 
	\begin{equation} 
		v_n^G = v_{n+1}^D + v_{n+1}^M. 
	\label{eq:Conservation_Law} 
	\end{equation}
 
	\noindent Both volumes contribute to the birth volume distribution for the next generation, $P_{n+1}(v)$.
	
	The question of stability then depends on the manner in which the cells control their growth. Building on the work of Refs. \cite{amir2014cell, ho2015simultaneous, tanouchi2015noisy}, we abstract away the myriad biological underpinnings of a cell's behavior (e.g. biochemical pathways, replication of DNA, etc.) by assuming a cell born with volume $v_n$ attempts to grow to its final (pre-division) size $v_n^G$ according to a species-specific growth policy, $v_n^G = f(v_n)$. Some of our results hold for arbitrary policies, but we focus mostly on the class of affine linear policies,
	
	\begin{equation}
		f(v) = \Delta + c \, v,
	\label{eq:Affine_Linear_Policy}
	\end{equation}
	
	\noindent in which the cell adds a constant volume $\Delta$ (independent of initial size) to its initial volume (weighted by a dimensionless control coefficient $c$) during growth. We should note this is a special case of a more general class of processes first considered from a thoroughly mathematical perspective in a series of papers (Refs. \cite{kesten1973random} and \cite{kesten1974renewal}) by Kesten in the 1970s. As is argued in Ref. \cite{amir2014cell}, we can always treat the affine linear policy as the first-order Taylor expansion of a nonlinear policy about the typical cell size. This procedure should result in a good approximation provided the noise is not too large. Note that the control coefficient $c$ in our model relates to the parameter $\alpha$ in Ref. \cite{amir2014cell} by 
	
	\begin{equation}
		c = 2(1-\alpha).
	\end{equation}
	
	This class of policies contains three special cases that have been proposed to describe the growth of certain organisms. The first case, known as the timer policy, is the policy with $\Delta = 0$,
	
	\begin{equation}
		f(v) = c \, v,
	\label{eq:Timer_Policy}
	\end{equation}
	
	\noindent which results in the cells growing for a constant time (see Eq. \ref{eq:Growth_Time}). However, as we show in Sec. \ref{Sec:Proof}, this policy results in an inherently unstable population. 
	
	The second case, the threshold or critical size policy \cite{donachie1968relationship}, arises in the limit as $c = 0$,

	\begin{equation}
		f(v) = \Delta.
	\label{eq:Threshold_Policy}
	\end{equation}
	
	\noindent As the name implies, under this policy the cells divide when they grow to a specific critical size $\Delta$, eliminating correlations between initial and final cell size and leading to a narrow stable distribution. However, this lack of correlation stands in contrast to experiment: studies in both bacteria \cite{koppes1980correlation} and yeast \cite{turner2012cell} suggest that final size is actually correlated with initial size.
	
	The third case, known as the incremental policy \cite{wang2010robust, amir2014cell, campos2014constant, soifer2016single}, sits between the threshold and timer policies,

	\begin{equation}
		f(v) = \Delta + v.
	\label{eq:Incremental_Policy}
	\end{equation}
	
	\noindent A cell following this policy attempts to increase its birth volume by a set amount $\Delta$. The molecular mechanism behind this policy is not fully understood \cite{robert2015size}. This policy is also appropriate in the case of damage control, as we show in the next section.
	
	Now to carry out any given policy, the cell must grow exponentially for a time $t^G(v_n)$ related to $f(v_n)$ by
	
	\begin{equation}
		f(v_n) = v_n \, e^{\lambda t^G(v_n)} \implies t^G(v_n) = \frac{1}{\lambda} \log\left( \frac{f(v_n)}{v_n} \right).
	\label{eq:Growth_Time}
	\end{equation}
	
	\noindent In reality, the cells cannot grow for precisely this time and so their actual fully-grown volume is affected by noise. The exponential dependence of $v_n^G$ on $t^G(v_n)$ (Eq. \ref{eq:Growth_Time}) then implies that errors in the growth time give rise to multiplicative noise,
	
	\begin{equation}
		v_n^G = v_n \, e^{\lambda (t_n^G + t_n^N)} = f(v_n) \, e^{\lambda t_n^N},
	\end{equation}
	
	\noindent where $t_n^N$ (superscript $N$ for ``noise'') is the growth timing error. We can also allow for additive noise, $v_n^N$, in which case our noisy growth model becomes
	
	\begin{equation}
		v_n^G = f(v_n) \, e^{\lambda t_n^N} + v_n^N.
	\label{eq:Noisy_Growth_Model}
	\end{equation}
	
	\noindent We should also note that, depending on how careful we are in choosing $f(v)$ and the distributions for the noise parameters, this model allows for cell volume to \emph{decrease}; however, given the low magnitude of noise we consider, the probability for this scenario is small enough to be negligible (see Table \ref{Table:Moment_Comparison} and Fig. \ref{fig:Stable_Distribution_Figure}).
	
	After the cell reaches its final volume $v_n^G$, we assume it divides such that $v_{n+1}^D = r \, v_{n+1}^M$, where the asymmetry ratio $0 \leq r \leq 1$ is always the same. Combining this assumption with volume conservation (Eq. \ref{eq:Conservation_Law}), we can write the post-division mother and daughter volumes directly in terms of the fully-grown cell volume:
	
	\begin{equation}
		v_{n+1}^M = \frac{1}{1+r} v_n^G, \qquad \qquad v_{n+1}^D = \frac{r}{1+r} v_n^G.
	\label{eq:Mother_Daughter_Volumes}
	\end{equation}

	\noindent Note that we can go from $v_{n+1}^M$ to $v_{n+1}^D$ by taking $r \to 1/r$, so in what follows it suffices to give the results for $v_{n+1}^M$, from which one can obtain the results for $v_{n+1}^D$ via this transformation. This symmetry under relabeling mothers and daughters must hold for all our population-level results.
	
	\subsection{Damage Control}
	
		We also consider how symmetrically-dividing cells can control their levels of damaged proteins by dividing these undesirable proteins asymmetrically between their daughter cells. Assuming the damaged proteins are not autocatalytic, we can describe the accrual of damaged proteins in a cell, $d$, by a volume-dependent rate $q\left( v(t) \right)$, such that
	
		\begin{equation}	
			\frac{d}{dt} \left( d(t) \right) = q\left( v(t) \right).
		\label{eq:Damage_Rate}
		\end{equation}
		
		\noindent Given that both the mechanisms that result in damaged proteins (e.g. environmental radiation or pollution) and the cell volume itself are stochastic, it follows that the rate of damage is also stochastic. Nevertheless, Eq. \ref{eq:Damage_Rate} still implies that the amount of damaged proteins in a cell at time $t$ after birth is given by
		
		\begin{equation}
			d(t) = d_0 + \exp\left[ \int_0^t d\tau \; q\left( v(\tau) \right) \right],
		\end{equation}
		
		\noindent where $d_0$ refers to the initial amount of damage present at birth. Thus it follows that the amount of damage an $n$th generation cell has at the time of division (the same time $t^G$ from Eq. \ref{eq:Growth_Time}) takes the same form as the incremental policy (Eq. \ref{eq:Incremental_Policy}),
	
		\begin{equation}
			d_n^G = \Delta + d_n + \delta\Delta,
		\label{eq:Damage_Model}
		\end{equation}
		
		\noindent where $d_n$ is the amount of damage an $n$th generation cell has at birth, $d_n^G$ is the final amount of damage, $\Delta$ is the average amount of damage added during the cycle,
		
		\begin{equation}
			\Delta = \left\langle \exp\left[ \int_0^{t^G} d\tau \; q\left( v(\tau) \right) \right] \right\rangle
		\end{equation}
		
		\noindent and $\delta\Delta$ is an additive (relative to the damage) noise term summarizing the stochasticity of the damage rate about the mean. Note that we have not yet made any assumptions as to the distribution for $\delta\Delta$. Furthermore, since we assume the cells divide their volume symmetrically, the asymmetry ratio $r$ and Eq. \ref{eq:Mother_Daughter_Volumes} (with $v$ replaced with $d$) now apply solely to the division of the damaged proteins.

\section{Deriving the Evolution Relation for the Cell Size Distribution}
\label{Sec:Model_Building}
	
	We begin our derivation of Eq. \ref{eq:Recursive_Relation} by calculating the distribution for $v_n^G$. Marginalizing over the birth volume $v_n$ and the noise parameters $t_n^N$ and $v_n^N$, we can write $P_n^G(v)$ as
	
	\begin{multline}
		P_n^G \left( v \right) = \int dv' P_n(v') \iint \, dt^N \, dv^N \; P\left( t^N, v^N \right) \\ \times P\left( v_n^G = v \, \middle| v_n = v', t^N, v^N \right),
	\label{eq:Formal_Expansion}
	\end{multline}
	\newline
	\noindent where the distribution $P\left(t^N, v^N \right)$ describes the noise, $P_n(v')$ describes the birth volume distribution for the $n$th generation and $P\left( v_n^G \middle| v_n, t_n^N, v_n^N \right)$ enforces the constraints of our noisy growth model (Eq. \ref{eq:Noisy_Growth_Model}),
	
	\begin{equation}	
		P\left( v_n^G \middle| v_n, t^N, v^N \right) = \delta\left( v_n^G - \left( f(v_n) \, e^{\lambda t^N} + v^N \right) \right).
	\end{equation}
	
	\noindent We assume the noise is independent across generations and independent of birth volume. Putting the above expression back into Eq. \ref{eq:Formal_Expansion}, we have
	
	\begin{widetext}
	\begin{equation}
		P_n^G\left( v \right) = \int dv' \; \left[ \iint dt^N \, dv^N \; \delta\left( v - \left( f(v') \, e^{\lambda t^N} + v^N \right) \right) P\left(t^N, v^N \right) \right] P_n\left( v' \right).
	\label{eq:Division_Volume}
	\end{equation}
	
	 Now the simple scaling relation between $v_{n+1}^M$ and $v_n^G$ (Eq. \ref{eq:Mother_Daughter_Volumes}) allows us to obtain the distribution for $v_{n+1}^M$ by a change of variables,
	
	\begin{equation}
		P_{n+1}^M \left( v \right) = (1+r) \, P_n^G\left[ (1+r) \, v \right],
	\end{equation}
	
	\noindent which we can write out fully as
	
	\begin{equation} \begin{split}
		P_{n+1}^M \left( v \right) &= (1+r) \int dv' \; \left[ \iint dt^N \, dv^N \; \delta\left( (1+r) \, v - \left( f(v') \, e^{\lambda t^N} + v^N \right) \right) P\left(t^N, v^N \right) \right] P_n\left( v' \right), \\
			&= \int dv' \; \left[ \iint dt^N \, dv^N \; \delta\left( v - \frac{1}{1+r} \left( f(v') \, e^{\lambda t^N} + v^N \right) \right) P\left(t^N, v^N \right) \right] P_n \left( v' \right).
	\label{eq:Post_Division_Distribution}
	\end{split} \end{equation}
	
	\noindent Since there are always as many mother cells as daughter cells in a given generation, our cell volume distribution for the next generation is given by:
	
	\begin{equation}
		P_{n+1}\left( v \right) = \frac{1}{2} \left[ P_{n+1}^M\left( v \right) + P_{n+1}^D \left(v \right) \right].
	\label{eq:Recombination_Equation}
	\end{equation}
	
	\noindent Using our expression for $P_{n+1}^M \left( v \right)$ from Eq. \ref{eq:Post_Division_Distribution} and then taking $r \to 1/r$ to get $P_{n+1}^D \left( v \right)$, we can rewrite Eq. \ref{eq:Recombination_Equation} as a recursive integral equation:
	
	\begin{multline}
		P_{n+1}\left( v \right) = \frac{1}{2} \int dv' \left( \iint dt^N \, dv^N \bigg[ \delta\left( v - \frac{1}{1+r} \left( f(v') \, e^{\lambda t^N} + v^N \right) \right) \right. \\
	 \left. + \delta\left( v - \frac{r}{1+r} \left( f(v') \, e^{\lambda t^N} + v^N \right) \right)  \bigg] P\left(t^N, v^N \right) \right) P_n \left( v' \right).
	\end{multline}
	
	\noindent This suggests that we should define the integral kernel $K(v, v')$ as
	
	\begin{equation}
		K(v, v') = \frac{1}{2} \iint dt^N \, dv^N \left[ \delta\left( v - \frac{1}{1+r} \left( f(v') \, e^{\lambda t^N} + v^N \right) \right) + \delta\left( v - \frac{r}{1+r} \left( f(v') \, e^{\lambda t^N} + v^N \right) \right)  \right] P\left(t^N, v^N \right),
	\label{eq:Kernel}
	\end{equation}
	
	\noindent which we can then use to write our evolution equation in the form of Eq. \ref{eq:Recursive_Relation}.
	
	\end{widetext}
	
	\phantom{digital gardening}
	\clearpage

\section{On the Existence and Uniqueness of the Stable Distribution}
\label{Sec:Proof}

	Having completed our derivation of the evolution equation for the cell volume distribution (Eq. \ref{eq:Recursive_Relation}), we are now in a position to study the properties of stable distributions (should they exist) by taking $P_{n+1}(v) = P_n(v) = P(v)$, in which case our recursive integral equation turns into a homogeneous Fredholm integral equation of the second kind (Eq. \ref{eq:Eigenvalue_Equation}). Evidently the stable distributions $P(v)$ are eigenfunctions of the linear operator $\int_0^\infty dv' \; K(v,v')$  with eigenvalue 1. However, finding analytic expressions for these eigenfunctions is difficult in practice: even if we were to consider only additive Gaussian noise, $v^N \sim \mathcal{N}(0,\sigma_v)$, the resulting form of the kernel (Eq. \ref{eq:Kernel}),
	
	\begin{widetext}
	\begin{equation}
			K(v,v') = \frac{1+r}{2\sqrt{2\pi} r \sigma_v} \left( r \, \exp\left[ - \frac{\left( v - \frac{1}{1+r} f(v') \right)^2}{2 \left(\sigma_v /(1+r) \right)^2} \right] + \exp\left[ - \frac{\left( v - \frac{r}{1+r} f(v') \right)^2}{2 \left( r \, \sigma_v / (1+r) \right)^2} \right] \right),
	\label{eq:Gaussian_Additive_Kernel}
	\end{equation}
	
	\noindent is complicated enough that an analytical solution is out of our reach. Though the fact that this kernel is the sum of two Gaussians might suggest that the stable distribution is also a mixture of two Gaussians, this is not the case (see Fig. \ref{fig:Stable_Damage_Distributions}). However, such a mixture is approximately valid for nearly-symmetric division. Our chances are no better when we trade additive noise for Gaussian multiplicative noise, $t^N \sim \mathcal{N}(0,\sigma_t)$, in which case the kernel,
	
	\begin{equation}
		K(v,v') = \frac{1}{2\sqrt{2\pi} \lambda \sigma_t} \frac{1}{v} \left( \exp\left[ - \frac{1}{2\lambda^2 \sigma_t^2} \left( \log\left[ \frac{v}{\frac{r}{1+r} f(v')} \right] \right)^2 \right] + \exp\left[ - \frac{1}{2\lambda^2 \sigma_t^2} \left( \log\left[ \frac{v}{\frac{1}{1+r} f(v')} \right] \right)^2 \right] \right),
	\label{eq:Gaussian_Multiplicative_Kernel}
	\end{equation}
	\end{widetext}
	
	\noindent becomes a mixture of log-normals instead of a mixture of Gaussians. Again, the immediate ansatz (a mixture of two log-normals) is not exactly a solution, but is approximately correct for $r \approx 1$ \cite{soifer2016single, amir2014cell}.
		
		Though casting the problem in terms of a homogeneous Fredholm integral equation of the second kind does not give us the ability to analytically construct stable distributions, it does enable us to apply powerful mathematical machinery in addressing questions of existence and uniqueness.
		
		For example, we can use functional analysis to help address the question of the existence of a stable distribution. Given our formal expression for the kernel (Eq. \ref{eq:Kernel}) in terms of Dirac delta functions, it is apparent that
		
		\begin{equation}
			\int_0^\infty dv \; K(v,v') = \int dt^N \, dv^N \; P(t^N, v^N) = 1.
		\label{eq:Eigenvalue_One}
		\end{equation}
		
		\noindent Hence the constant function $q(v) = 1$ is a left eigenfunction of our kernel with eigenvalue $1$. Thus we know that $1$ is an eigenvalue of the adjoint of the kernel, which implies that it is also an eigenvalue of the kernel itself, and so we know that our linear operator admits at least one right eigenfunction with eigenvalue 1.
		
		Now while it is true that Eq. \ref{eq:Eigenvalue_Equation} admits at least one solution, it could be the case that none of these solutions can actually be considered a \emph{probability distribution}. There are two possible issues: first, there may not be any non-negative eigenfunctions  (i.e. eigenfunctions $q(v)$ such that $q(v) \geq 0$ $\forall$ $v$) and second, should a non-negative eigenfunction exist, it may not be normalizable in the $L^1$ sense. The issue of normalizability arises due to the fact that the stable distribution $P(v)$ is defined on the infinite domain $[0,\infty)$; the problem vanishes in the discrete case (finding the stable distribution for a Markov chain).
		
		The problem of normalizability does in fact occur under the timer model (Eq. \ref{eq:Timer_Policy}), under which the growth time is independent of the initial cell size, $t^G = \log(c) / \lambda$ (Eq. \ref{eq:Growth_Time}), and there are no natural volume scales in the system. Since every cell grows exponentially for a time $t^G$, we expect cell size to perform a geometric random walk, falling off as $1 / v$ (which is non-normalizable). We can check this ansatz directly. First we note that the kernel (Eq. \ref{eq:Gaussian_Multiplicative_Kernel}) becomes

		\begin{widetext}
		\begin{equation}
			K(v,v') = \frac{1}{2\sqrt{2\pi} \lambda \sigma_t} \frac{1}{v} \left( \exp\left[ - \frac{\left( \log[v'] - \log\left[ \frac{1+r}{c r} v \right] \right)^2}{2\lambda^2 \sigma_t^2} \right] + \exp\left[ - \frac{\left( \log[v'] - \log\left[ \frac{1+r}{c} v \right] \right)^2}{2\lambda^2 \sigma_t^2} \right] \right).
		\end{equation}
		\end{widetext}
		
		\phantom{limb}
		\clearpage
		
		\noindent Multiplying this kernel by a factor of $1 / v'$, we can rewrite the resulting expression as the sum of two log-normal distributions in $v'$,
		
		\begin{multline}
			 K(v,v') \frac{1}{v'} = \frac{1}{2v} \left[ \log\,\mathcal{N}_{v'} \! \! \left( \log\left[\frac{1+r}{c r} v \right] , \; \lambda^2 \sigma_t^2 \right) \right. \\ \left. + \log\,\mathcal{N}_{v'} \! \! \left( \log\left[\frac{1+r}{c} v \right], \; \lambda^2 \sigma_t^2 \right) \right].
		\end{multline}
		
		\noindent If we then integrate over $v'$, the two lognormal distributions integrate to 1, leaving us with
		
		\begin{equation}
			\int_0^\infty dv' \; K(v,v') \, \frac{1}{v'} = \frac{1 + 1}{2v} = \frac{1}{v},
		\end{equation}
		
		\noindent and so the non-normalizable function $1 / v'$ is indeed a right eigenfunction with eigenvalue 1. Moreover, since the only parameter in the system with units of volume is $v$ itself, it follows from dimensional analysis that this is the \emph{unique} eigenfunction. Thus the timer model is inherently unstable.
		
	However, when a stable distribution does exist, our mathematical machinery helps us show that it is unique. Assuming there exists a finite (nonzero) number of probability distributions satisfying Eq. \ref{eq:Eigenvalue_Equation}, the normalizability ($\int dv \; P(v) = 1$) and positivity ($P(v) > 0$) of the distributions imply that we can always choose a cutoff volume $\mathcal{V}$ such that $P(v > \mathcal{V}) < \epsilon$ for every distribution for any $\epsilon > 0$. Hence we can choose $\epsilon$ small enough to safely discretize Eq. \ref{eq:Eigenvalue_Equation} (a process we describe in detail in App. \ref{App:Discretization}). More explicitly, we convert our integral equation into a finite-dimensional matrix equation,
		
		\begin{equation}
			\pvec_{n+1} = K \, \pvec_n,
		\label{eq:Discrete_Eigenvalue_Equation}
		\end{equation}
		
		\noindent where $\pvec_n$ is a discrete probability vector and $K \in \mathbb{R}^{(N+1) \times (N+1)}$ is a discrete version of Eq. \ref{eq:Kernel}. As we show in App. \ref{App:Discretization}, we can construct $K$ such that it is a non-negative, connected stochastic matrix. Thus it follows from the Perron-Frobenius theorem that 1 is the largest eigenvalue of $K$ and its corresponding eigenvector is unique and non-negative. Moreover, since 1 is the largest eigenvalue of $K$, the system is guaranteed to converge to a unique, stable distribution over successive generations.
		
		While it may appear that a similar argument would also prove the \emph{existence} of the stable distribution, the situation is more nuanced. In order for Eq. \ref{eq:Discrete_Eigenvalue_Equation} to be a good approximation of the continuous case, the probability for a cell to be in the overflow bin ($v > \mathcal{V}$), i.e. the last element of $\pvec$, must decay to zero as $\mathcal{V} \to \infty$. However, this is not \emph{a priori} guaranteed; indeed, for unstable systems, the probability for being in the last bin remains finite even as we increase $\mathcal{V}$, whereas in stable systems we do in fact see the overflow probability decay with $\mathcal{V}$. Thus by checking the cutoff dependence of the overflow probability in the discretized problem, we can attempt to numerically determine whether the continuous problem (Eq. \ref{eq:Eigenvalue_Equation}) admits a stable distribution, as we illustrate in Fig. \ref{fig:Overflow_Probability_Figure} for both stable and unstable systems.
	
	Though the numerical method for addressing the question of existence can be useful, we would prefer to formally answer the question with functional analysis. Unfortunately, the infinite domain over which the distribution is defined complicates the standard approaches. We instead provide an alternative avenue by which to analyze the stability of the system. Beginning in Sec. \ref{Sec:Moments}, we carry out a moment-based analysis that elucidates the statistical properties of stable distributions. Then, motivated by the manner in which certain moments cease to exist (Eq. \ref{eq:Moment_Constraint}), we seek self-consistent power-law-tailed solutions to Eq. \ref{eq:Eigenvalue_Equation} and use the power of the tail to asymptotically determine when the solution is normalizable (see Sec. \ref{Sec:Tail}). This asymptotic analysis, taken together with the numerical approach, allows us to find conditions for the existence of the stable distribution and argue for its uniqueness, non-negativity and normalizability. 
		

\section{Statistical properties of the stable distribution under affine linear growth policies}
\label{Sec:Moments}

	Here we assume a unique stable distribution exists and we focus on elucidating its statistical properties, namely its moments. Since the stable distribution is related to the mother and daughter distributions by Eq. \ref{eq:Recombination_Equation}, we can calculate the $k$th moment as
	
	\begin{equation} \begin{split}
		\langle v^k \rangle &= \frac{1}{2} \int_0^\infty dv \; v^k \left( P^M(v) + P^D(v)  \right), \\
			&= \frac{1}{2} \left( \left\langle ( v^M )^k \right\rangle + \left\langle ( v^D )^k \right\rangle \right),
	\end{split} \end{equation}
	
	\noindent where we can use Eq. \ref{eq:Mother_Daughter_Volumes} to write the moments $\left\langle ( v^M )^k \right\rangle$ and $\left\langle ( v^D )^k \right\rangle$ in terms of $\left\langle (v^G)^k \right\rangle$,
	
	\begin{equation} \begin{split}
		\left\langle (v^M)^k \right\rangle &= \frac{1}{(1+r)^k} \left\langle (v^G)^k \right\rangle, \\
		\left\langle (v^D)^k \right\rangle &= \frac{r^k}{(1+r)^k} \left\langle (v^G)^k \right\rangle.
	\end{split} \end{equation}
	
	\noindent Using our noisy growth model (Eq. \ref{eq:Noisy_Growth_Model}) to write $\left\langle (v^G)^k \right\rangle$ in terms of the birth volume $v$,
	
	\begin{equation}
		\left\langle ( v^G )^k \right\rangle = \left\langle \left( f(v) \, e^{\lambda t^N} + v^N \right)^k \right\rangle,
	\label{eq:Kth_Moment_Growth}
	\end{equation}
	
	\noindent we then obtain an equation for the $k$th moment of $v$ which itself depends on moments of $v$,
	
	\begin{equation}
		\langle v^k \rangle =  \frac{1+r^k}{2(1+r)^k} \, \left\langle \left( f(v) \, e^{\lambda t^N} + v^N \right)^k \right\rangle.
	\label{eq:Kth_Moment}
	\end{equation}

	Setting $k = 1$ in Eq. \ref{eq:Kth_Moment}, we find that the mean of the stable distribution satisfies the following relation,
	
	\begin{equation}
		\langle v \rangle = \frac{1}{2} \, \left\langle f(v) \, e^{\lambda t^N} + v^N \right\rangle = \frac{1}{2} \left \langle f(v) \right\rangle \left\langle e^{\lambda t^N} \right \rangle
	\label{eq:Proto_Mean}
	\end{equation}

	\noindent where we assume that the expected value of the additive noise is zero, $\langle v^N \rangle = 0$, and that the birth volume and multiplicative noise are independent. We have not yet made any assumptions on the multiplicative noise; however it is useful to note that, for Gaussian $t^N$ (i.e. $t^N \sim \mathcal{N}(0,\sigma_t))$, 
	
	\begin{equation}
		\left\langle e^{k \lambda t^N} \right\rangle = e^{\frac{1}{2} k^2 \lambda^2 \sigma_t^2}.
	\label{eq:Multiplicative_Expectation}
	\end{equation}
	
	\noindent For the sake of generality, we continue using the more general moments.
	
	Now were it the case that $\langle f(v) \rangle = f(\langle v \rangle)$, then Eq. \ref{eq:Proto_Mean} would furnish a possibly nonlinear consistency equation which we could solve for $\langle v \rangle$. However, this is generally not true: the exception being when $f(v)$ is an affine linear policy (Eq. \ref{eq:Affine_Linear_Policy}). Assuming the cells follow such a policy, Eq. \ref{eq:Proto_Mean} becomes an explicit equation for $\langle v \rangle$,

	\begin{equation}
		\langle v \rangle = \frac{1}{2} \left(\Delta + c \, \langle v \rangle \right) \left\langle e^{\lambda t^N} \right \rangle,
	\end{equation}
	
	\noindent implying that the mean of the stable distribution is

	\begin{equation}
		\langle v \rangle = \frac{\Delta \langle e^{\lambda t^N} \rangle}{2 - c \langle e^{\lambda t^N} \rangle}.
	\label{eq:Stable_Mean}
	\end{equation}
	
	\noindent As for the variance of the stable volume distribution, we need to know the second moment $\langle v^2 \rangle$, which we can compute in a similar manner:
	
	\begin{widetext}
	\begin{equation}
		\langle v^2 \rangle = \frac{\Delta^2 \langle e^{2\lambda t^N} \rangle}{ 2 \frac{(1+r)^2}{1+r^2} - c^2 \langle e^{2\lambda t^N} \rangle} \left( \frac{\left\langle (v^N)^2 \right\rangle}{\Delta^2 \langle e^{2\lambda t^N} \rangle} + \frac{2 + c \langle e^{\lambda t^N} \rangle}{2 - c \langle e^{\lambda t^N} \rangle} \right),
	\label{eq:Stable_Second_Moment}
	\end{equation}
	
	\noindent which leads to the following expression for the variance:
	
	\begin{equation}
		\sigma_v^2 = \frac{\Delta^2 \langle e^{2\lambda t^N} \rangle}{ 2 \frac{(1+r)^2}{1+r^2} - c^2 \langle e^{2\lambda t^N} \rangle} \left( \frac{\left\langle (v^N)^2 \right\rangle}{\Delta^2 \langle e^{2\lambda t^N} \rangle} + \frac{4 \langle e^{2\lambda t^N} \rangle - 2 \frac{(1+r)^2}{1+r^2} \langle e^{\lambda t^N} \rangle^2}{(2 - c \langle e^{\lambda t^N} \rangle)^2 \, \langle e^{2\lambda t^N} \rangle} \right).
	\label{eq:Stable_Variance}
	\end{equation}
	\end{widetext}
	
\begin{figure}[t]
		\centering
			\includegraphics[width=\linewidth]{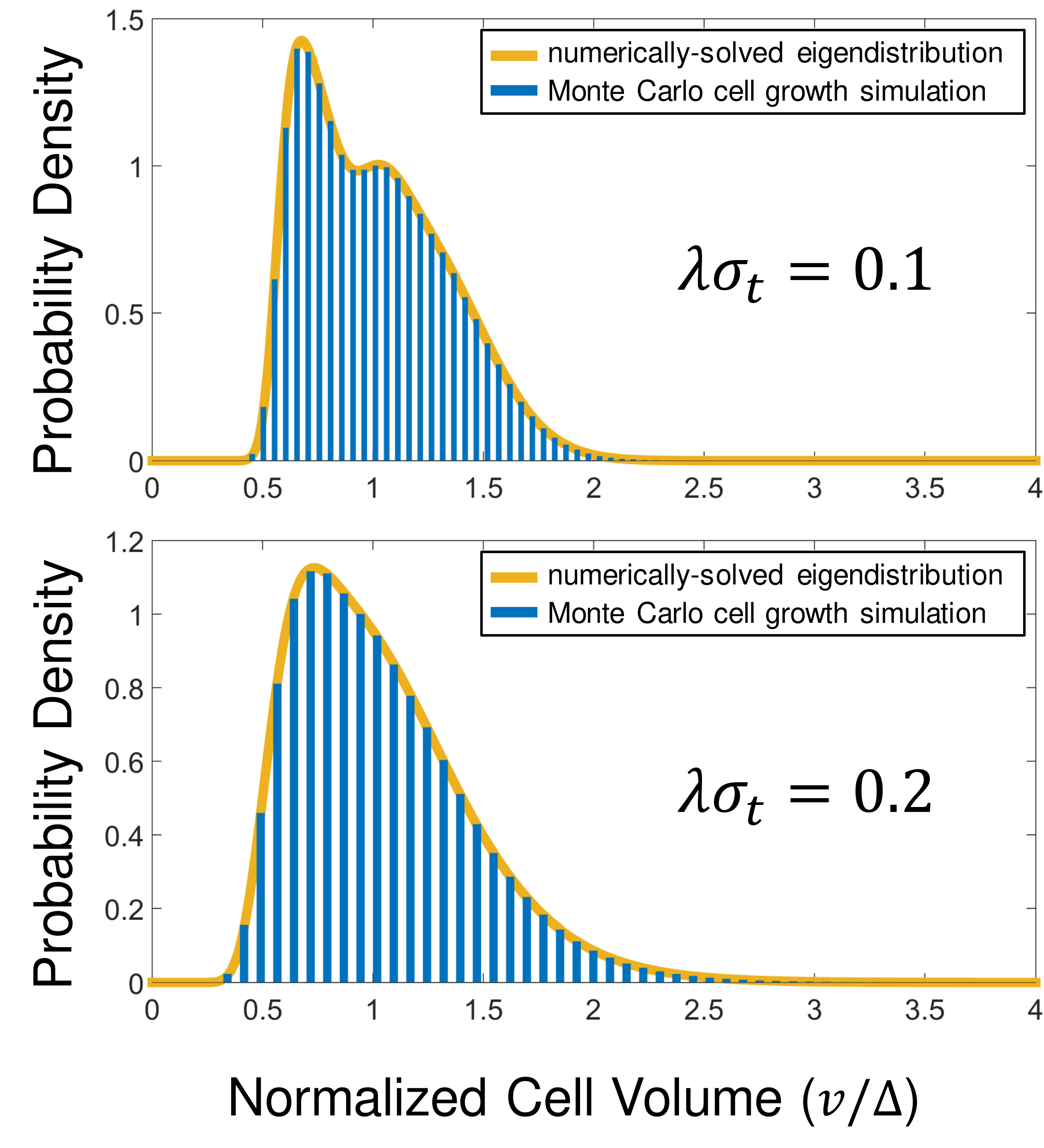}
		\caption{(Color online) Stable cell volume distribution for an affine linear model $f(v) = \Delta + c \, v$ with control coefficient $c = 1$ and asymmetry ratio $r = 0.6$ under two different multiplicative noise amplitudes: $\lambda \sigma_t = 0.1$ (top) and $\lambda \sigma_t = 0.2$ (bottom). The blue bars represent the results of a direct simulation of asymmetric cell growth (see Table \ref{Table:Moment_Comparison} for more details). The stable solution can be bimodal, as is the case for $\lambda\sigma_t = 0.1$. The results of these simulations compare well with the numerically-solved eigendistributions of our eigenvalue equation (\ref{eq:Eigenvalue_Equation}) which we obtained using a Fredholm integral equation solver by K. Atkinson and L. Shampine \cite{atkinson2008algorithm}.}
		\label{fig:Stable_Distribution_Figure}
	\end{figure}

	Our formulae for the mean and variance (Eqs. \ref{eq:Stable_Mean} and \ref{eq:Stable_Variance}) agree well with simulations (Table \ref{Table:Moment_Comparison}). We also employ a Fredholm integral equation solver by K. Atkinson and L. Shampine \cite{atkinson2008algorithm} to numerically solve Eq. \ref{eq:Eigenvalue_Equation} for the stable distribution (using the multiplicative-noise kernel from Eq. \ref{eq:Gaussian_Multiplicative_Kernel}) and compare with direct cell growth simulations in Fig. \ref{fig:Stable_Distribution_Figure}. The two methods agree well with each other and offer different computational advantages: the Monte Carlo approach is easier to implement than an integral equation solver, but takes longer to converge to the stable distribution than it takes to numerically solve the integral equation.
	
	These exact expressions for the mean and variance of the stable distribution are not only experimentally useful, but also of great theoretical interest. Cell volumes are always positive and so all moments of the stable distribution should also be positive. However, Eq. \ref{eq:Stable_Mean} implies the first moment is only positive when
	
	\begin{equation}
		c < \frac{2}{\langle e^{\lambda t^N} \rangle} = c^{(1)}_\textrm{max}.
	\label{eq:Mean_Constraint}
	\end{equation}
	
	\noindent Likewise, the second moment (Eq. \ref{eq:Stable_Second_Moment}) and variance (Eq. \ref{eq:Stable_Variance}) are positive only when the more stringent condition

	\begin{equation}
		c < \frac{\sqrt{2} \, (1+r)}{\sqrt{ (1 + r^2) \langle e^{2\lambda t^N} \rangle }} = c^{(2)}_\textrm{max},
	\label{eq:Variance_Constraint}
	\end{equation}
	
	\noindent is met. What then does it mean when the control coefficient $c$ is large enough that our formal expressions predict negative moments? This does not necessarily imply that the stable \emph{distribution} ceases to exist: rather, it shows that the tail of the distribution becomes heavy enough that these \emph{moments} cease to exist. Probability distributions for which some or all moments do not exist (e.g. the Cauchy distribution), despite being labeled, on occasion, as pathological, are still valid probability distributions. 

	These moment constraints suggest that we might be able to analyze the behavior of the tail of the stable distribution by understanding which of its moments do and do not exist. This of course requires us to derive an expression for the $k$th moment, a calculation we leave for App. \ref{App:kth_Moment}. Once again, since all the moments of the volume distribution should be positive, it follows from Eq. \ref{eq:Explicit_Kth_Moment} that the $k$th moment only exists if
		
		\begin{equation}
			2(1+r)^k > \left( 1 + r^k \right) c^k \left\langle e^{k \lambda t^N} \right\rangle,
		\end{equation}
		
		\noindent which implies that the $k$th moment of the stable distribution exists only if the control coefficient $c$ satisfies
		
		\begin{equation}
			c < (1+r) \left( \frac{2}{\left( 1 + r^k \right) \left\langle e^{k \lambda t^N} \right\rangle } \right)^{\frac{1}{k}} = c^{(k)}_\textrm{max}.
		\label{eq:Moment_Constraint}
		\end{equation}
		
		Interestingly, the properties of the additive noise do not appear to factor into this constraint at all. Nevertheless, Eq. \ref{eq:Moment_Constraint} still holds when we have only additive noise, as we explore in Sec. \ref{Sec:Damage_Control}. To be more concrete, we consider the case of Gaussian multiplicative noise $t^N \sim \mathcal{N}(0,\sigma_t)$ (and optional additive noise), in which case this constraint becomes
		
		\begin{equation}
			c < (1+r) \, e^{-\frac{1}{2} k \lambda^2 \sigma_t^2} \left( \frac{2}{ 1 +  r^k} \right)^{\frac{1}{k}}.
		\end{equation}
		
		\noindent Notice that, for $\sigma_t \neq 0$, the right hand side is a monotonically decreasing function of $k$ and as we take $k \to \infty$, the constraint approaches $c = 0$, implying that if the $k$th moment exists, so too do all the lower moments $(k-1), (k-2), \ldots, 0$. Furthermore, so long as multiplicative noise is present and $c > 0$, there will always be an integer $k^*$ past which the moments cease to exist (the case for solely additive noise is more nuanced, see Sec. \ref{Sec:Damage_Control}). This suggests that the stable distribution has a power law tail $P(v) \sim 1 / v^{1+\beta}$ with $k^* < \beta \leq k^* + 1$, which one can see more formally in Ref. \cite{kesten1973random}.  Note that this power-law tail is not captured by the methods used in Refs. \cite{amir2014cell} and \cite{brenner2015universal}, in which the stable distribution is approximated as log-normal in form.
		
		\begin{figure}
			\centering
				\includegraphics[width=\linewidth]{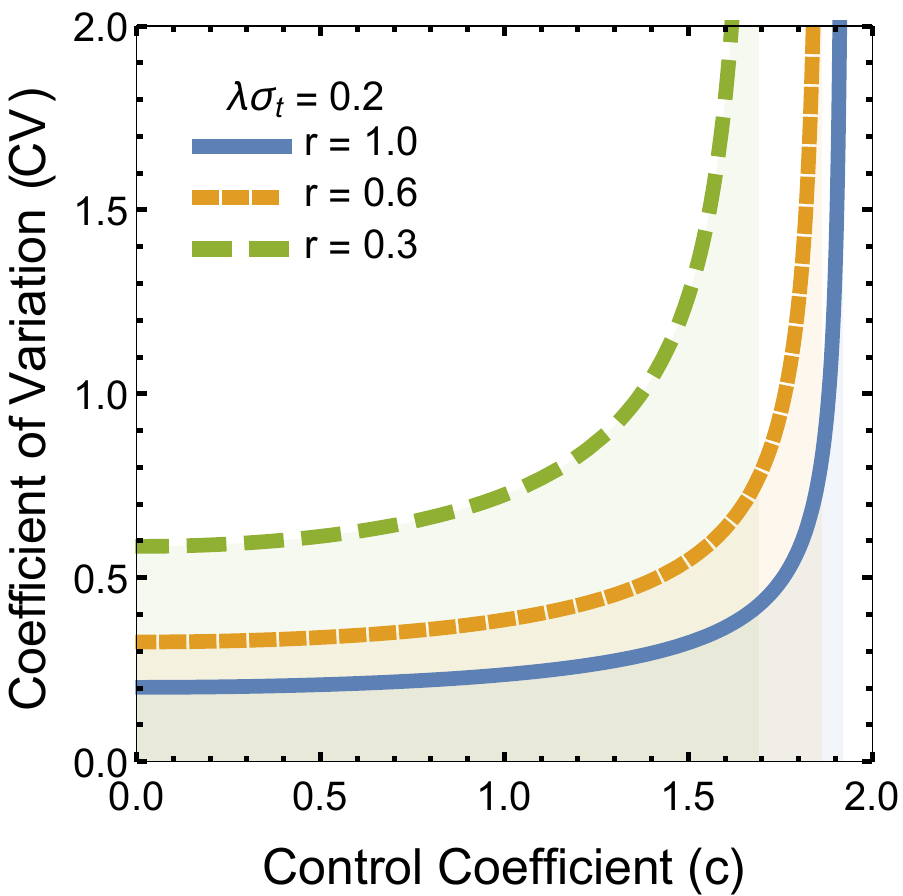}
			\caption{(Color online) Dependence of the coefficient of variation (CV) on the control coefficient $c$ for three asymmetry ratios $r = \{1.0, 0.6, 0.3\}$ with multiplicative noise of amplitude $\lambda\sigma_t = 0.2$. We give the analytical expression for the CV in Eq. \ref{eq:Inferring_CV}. Note that the value of $c$ at which the CV diverges (which we denote by $\cmax^{(2)}$ in Eq. \ref{eq:Variance_Constraint}), decreases with increasing asymmetry.}
		\label{fig:CV_Plot}
		\end{figure}
		
		We can also use the bounds on the control coefficient to rewrite the first and second moments more compactly, 
	
	\begin{equation} \begin{split}
		\langle v \rangle &= \frac{\Delta}{\cmax^{(1)} - c}, \\
		\langle v^2 \rangle &= \frac{\Delta^2}{(\cmax^{(2)})^2 - c^2} \left( \frac{\cmax^{(1)} + c}{\cmax^{(1)} - c} \right).
	\end{split} \end{equation}
	
	\noindent These expressions also lead to a simple and elegant relation for the coefficient of variation (CV),

	\begin{equation} \begin{split}
		\textrm{CV}^2 &= \frac{\sigma_v^2}{\langle v \rangle^2} = \frac{\langle v^2 \rangle - \langle v \rangle^2}{\langle v \rangle^2}, \\
			&= \frac{(\cmax^{(1)})^2 - (\cmax^{(2)})^2}{ (\cmax^{(2)})^2 - c^2 },
	\label{eq:CV}
	\end{split} \end{equation}
	
	\noindent from which it is apparent that $\textrm{CV} \in [\cvmin, \infty)$, where the minimum value, $\cvmin$, occurs at $c = 0$,
	
	\begin{equation}
		\cvmin^2 = \left( \frac{\cmax^{(1)}}{\cmax^{(2)}} \right)^2 - 1.
	\end{equation}
	
	\noindent Thus we can write the CV as

	\begin{equation}
		\textrm{CV} = \cvmin \, \left( 1 - \left( \frac{c}{\cmax^{(2)}} \right)^2 \right)^{-1/2}.
	\label{eq:Inferring_CV}
	\end{equation}
	
	\noindent As expected, turning up the control coefficient also increases the coefficient of variation. Of course as $c \to \cmax^{(2)}$, the CV diverges along with the second moment. We plot the dependence of the CV on $c$ and $r$ for multiplicative noise with $\lambda\sigma_t = 0.2$ in Fig. \ref{fig:CV_Plot}.


\section{An Asymptotic Analysis of the Tail of the Stable Distribution}
\label{Sec:Tail}

	\begin{figure}
		\centering
			\includegraphics[width=\linewidth]{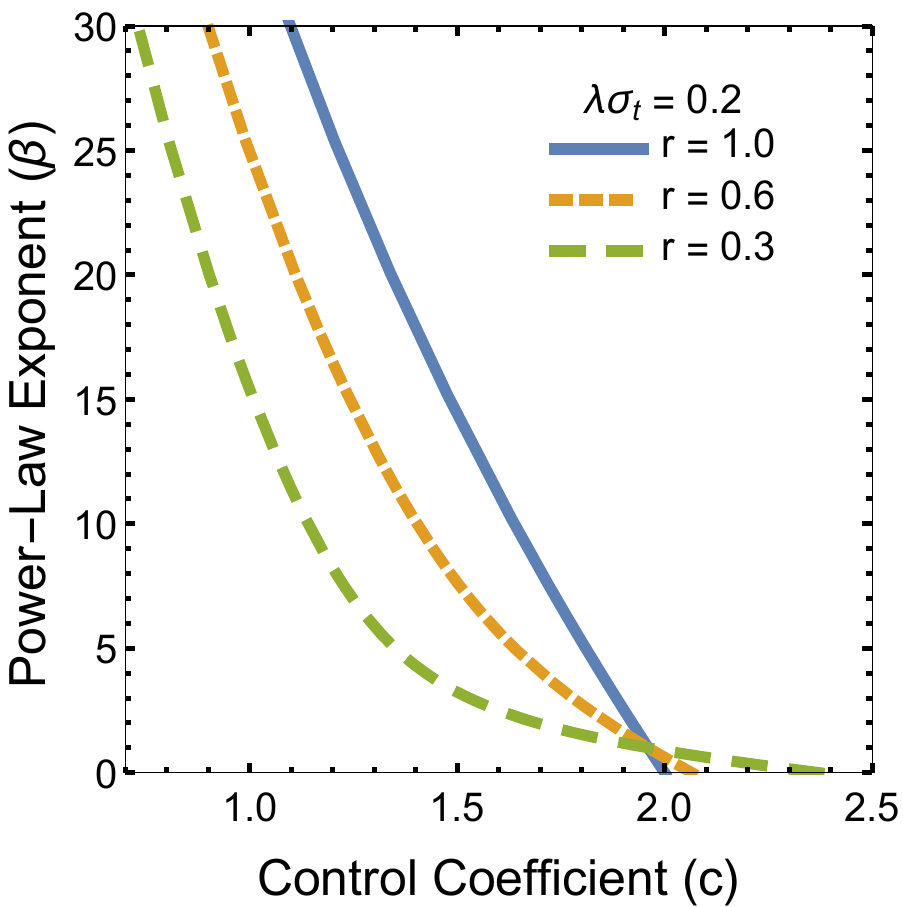}
		\caption{(Color online) Dependence of the power law exponent $\beta$ on the linear control coefficient $c$ for asymmetry ratios $r = \{1.0, 6.0, 3.0\}$ and multiplicative noise amplitude $\lambda \sigma_t = 0.2$. These curves all satisfy our consistency equation (\ref{eq:Power_Law_Exponent_Constraint}). As we increase $\lambda\sigma_t$ we also increase the spread (variance) of the stable volume distribution and induce a heavier tail. In the symmetric case, $\beta = 0$ at $c = 2$, indicating that the system becomes unstable. Interestingly, for asymmetric division, $\beta = 0$ for $c > 0$, implying that asymmetry can actually act to stabilize the system while at the same time bringing about a heavier tail (see Fig. \ref{fig:Phase_Diagrams}).}
		\label{fig:Power_Law_Exponent_Figure}
	\end{figure}

	Our work in Sec. \ref{Sec:Moments} suggests that the stable cell volume distribution under an affine linear policy possesses a power-law tail. However, our previous analysis only allows us to bound the power $\beta$ of this tail between $k^*$ (the power of the highest extant moment) and $k^* + 1$, i.e. $k^* < \beta \leq k^* + 1$. Here we carry out an asymptotic analysis of the tail to extend our results to non-integer powers and to study when and how an initial distribution converges to a stable distribution. Our analysis also allows us to study the stability properties of a larger class of policies,
	
	\begin{equation}
		f(v) = \Delta + c \, \Delta \left( \frac{v}{\Delta} \right)^\alpha,
	\label{eq:Generalized_Policy}
	\end{equation}
	
	\noindent where we now allow for a nonlinear dependence on $v$ controlled by an exponent $\alpha$. Note that (\ref{eq:Generalized_Policy}) reduces to the affine linear policy (\ref{eq:Affine_Linear_Policy}) when $\alpha = 1$. 
	
	Our asymptotic analysis proceeds by considering how an initial distribution with a power law tail evolves over successive generations under Eq. \ref{eq:Recursive_Relation}. Thus we assume our initial distribution $P_0(v)$ behaves as a power law with exponent $1 + \beta$ when $v \gg V_0$ for some cutoff volume $V_0$,
	
	\begin{equation}
		P_0(v) = \frac{\mathcal{N}}{v^{1+\beta}} \quad \textrm{ for } \quad v \gg V_0,
	\label{eq:Initial_Distribution}
	\end{equation}
	
	\noindent where $\mathcal{N}$ is a normalization constant. The distribution for the next generation, $P_1(v)$, is then given by Eq. \ref{eq:Recursive_Relation},
	
	\begin{equation}
		P_1(v) = \int_0^\infty dv' \; K(v,v') \, P_0(v').
	\label{eq:First_Recursion}
	\end{equation}
	
	\noindent Since we only wish to specify the tail of $P_0(v)$, it will be useful to split the integral in Eq. \ref{eq:First_Recursion} into two parts,
	
	\begin{multline}
		P_1(v) = \int_0^{V_0} dv' \; K(v,v') \, P_0(v') \\ + \int_{V_0}^\infty dv' \; K(v,v') \, \frac{\mathcal{N}}{(v')^{1+\beta}},
	\label{eq:Split_Asymptotic}
	\end{multline}
	
	\noindent where we have inserted the power law form of $P_0(v')$ (Eq. \ref{eq:Initial_Distribution}) in the integral over $v' \in [V_0, \infty)$. In order for our asymptotic analysis to work, we must both evaluate the second term and show that the first term, for which we do not know the exact form of $P_0(v')$, falls off with $v$ faster than a power law tail. 
	
	In order to make progress, we assume that we have Gaussian multiplicative noise and no additive noise, allowing us to use the kernel from Eq. \ref{eq:Gaussian_Multiplicative_Kernel}. For computational convenience, we split this kernel into two terms,
	
	\begin{equation}
		K(v,v') = \frac{1}{2} \left( K\left(v,v';r\right) + K\left(v,v';\frac{1}{r} \right) \right),
	\end{equation}
	
	\noindent where we define the subkernel $K(v,v';r)$ by
	
	\begin{multline}
		K(v,v';r) \\ = \frac{1}{\sqrt{2\pi} \lambda \sigma_t \, v} \exp\left[ - \frac{1}{2\lambda^2 \sigma_t^2} \left( \log\left[ \frac{(1+r)v}{r f(v')} \right] \right)^2 \right].
	\label{eq:Subkernel_Full}
	\end{multline}
	
	Let us now consider the first term in Eq. \ref{eq:Split_Asymptotic}. The mean value theorem implies there exists a $\hat{v}_0 \in [0, V_0]$ such that
	
	\begin{equation} \begin{split}
		\int_0^{V_0} dv' \; K(v,v') \, P_0(v') &= K(v, \hat{v}_0) \int_0^{V_0} dv' \; P_0(v'), \\ 
		&\leq K(v, \hat{v}_0),
	\end{split} \end{equation}
	
	\noindent where we use the normalization ($\int_0^\infty dv' \; P_0(v') = 1$) and positivity ($P_0(v') \geq 0 \; \forall \; v' \in [0,\infty))$ of the probability distribution to bound this first term from above by $K(v,\hat{v}_0)$. Now since $K(v,\hat{v}_0)$ is log-normal in $v$, it follows that this term falls off with $v$ no slower than
	
	\begin{equation}
		\int_0^{V_0} dv' \; K(v,v') \, P_0(v') \sim \mathcal{O}\left( \frac{1}{v^{1+\log{v}}} \right).
	\end{equation}
	
	\noindent Since we only care about the tail of $P_1(v)$, we can always take $v$ large enough to make this term fall off faster than any fixed-power tail, and so we can safely ignore the contribution of the first term in Eq. \ref{eq:Split_Asymptotic}.
	
	Turning our attention to the second term in Eq. \ref{eq:Split_Asymptotic}, we first split it into two pieces,
	
	\begin{equation}
		\mathcal{N} \int_{V_0}^\infty dv' \; \frac{K(v,v')}{(v')^{1+\beta}} = \frac{\mathcal{N}}{2} \left( K(v;r) + K\left(v; \frac{1}{r} \right) \right),
	\label{eq:Two_Pieces}
	\end{equation}
	
	\noindent where we define the integrated subkernel $K(v;r)$ by
	
	\begin{equation}
		K(v;r) = \int_{V_0}^\infty dv' \; K(v,v';r) \frac{1}{(v')^{1+\beta}},
	\end{equation}
		
	\noindent and $K(v;1/r)$ is obtained by taking $r \to 1/r$ in $K(v;r)$. Unfortunately the form of our generalized growth policy (Eq. \ref{eq:Generalized_Policy}) makes it rather difficult to perform these integrals. However, since the integral runs over $v' \in [V_0, \infty)$ and $V_0 \gg \Delta$, we can approximate our generalized growth policy (Eq. \ref{eq:Generalized_Policy}) as
	
	\begin{equation}
		f(v') \approx c \, \Delta \left( \frac{v'}{\Delta} \right)^\alpha,
	\end{equation}
	
	\noindent and thereby approximate the subkernel $K(v,v';r)$ by
	
	\begin{widetext}
	\begin{equation}
		K(v,v';r) \approx \frac{1}{\sqrt{2\pi} \lambda \sigma_t \, v} \exp\left[ - \frac{1}{2\lambda^2 \sigma_t^2} \left( \log\left[ \frac{c \, r}{1+r} \frac{\Delta}{v} \left( \frac{v'}{\Delta} \right)^\alpha \right] \right)^2 \right].
	\label{eq:Subkernel}
	\end{equation}
	
	\noindent Thus the integrated subkernel has the form
	
	\begin{equation}
		K(v;r) \approx \frac{1}{v} \int_{V_0}^\infty \frac{dv'}{\sqrt{2\pi} \lambda \sigma_t \, v'} \frac{1}{(v')^\beta} \exp\left[ - \frac{1}{2\lambda^2 \sigma_t^2} \left( \log\left[ \frac{c \, r}{1+r} \frac{\Delta}{v} \left( \frac{v'}{\Delta} \right)^\alpha \right] \right)^2 \right].
	\end{equation}
	
	\noindent We can evaluate this integral exactly by changing variables from $v'$ to $x$, defined as
	
	\begin{equation}
		x = \frac{v'}{\Delta} \left( \frac{c \, r}{1+r} \frac{\Delta}{v} \right)^{\frac{1}{\alpha}},
	\end{equation}
	
	\noindent which allows us to write $K(v;r)$ as
	
	\begin{equation}
		K(v;r) \approx \frac{ \Delta^{\beta\left(\frac{1}{\alpha}-1 \right)} }{ \alpha \, v^{1+\frac{\beta}{\alpha}}} \left( \frac{c \, r}{1+r} \right)^{\frac{\beta}{\alpha}} \int_{x_0}^\infty dx \; x^{-\beta} \left( \frac{\alpha}{\sqrt{2\pi} \lambda \sigma_t \, x } \exp\left[ - \frac{\alpha^2 \log^2 x}{2\lambda^2 \sigma_t^2} \right] \right),
	\end{equation}
	
	\noindent where $x_0$ is a function of $v$,
	
	\begin{equation}
		x_0 = \frac{V_0}{\Delta} \left( \frac{c \, r}{1+r} \frac{\Delta}{v} \right)^{\frac{1}{\alpha}}
	\end{equation}
	
	In the large $v$ limit, i.e. $v \gg V_0$, the lower limit of integration approaches zero. Thus this integral reduces to the expectation value of $x^{-\beta}$ for a log-normal distribution with $\sigma = \lambda \sigma_t / \alpha$:
	
	\begin{equation}
		\int_0^\infty dx \; x^{-\beta} \left( \frac{\alpha}{\sqrt{2\pi} \lambda \sigma_t \, x } \exp\left[ - \frac{\alpha^2 \log^2 x}{2\lambda^2 \sigma_t^2} \right] \right) = \exp\left[ \frac{\beta^2 \lambda^2 \sigma_t^2}{2\alpha^2} \right].
	\end{equation}
	\end{widetext}
	
	\phantom{menace}
	\clearpage
	\noindent Using this, we can write $K(v;r)$ as
	
	\begin{equation}
		K(v;r) \approx \frac{ \Delta^{\beta\left(\frac{1}{\alpha}-1 \right)} }{ \alpha \, v^{1+\frac{\beta}{\alpha}}} \left( \frac{c \, r}{1+r} \right)^{\frac{\beta}{\alpha}} \exp\left[ \frac{\beta^2 \lambda^2 \sigma_t^2}{2\alpha^2} \right],
	\end{equation}
	
	\noindent from which we can obtain $K(v;1/r)$ by taking $r \to 1/r$. Substituting $K(v;r)$ and $K(v; 1/r)$ into Eq. \ref{eq:Two_Pieces} then yields, for $v \gg V_0$,
	
	\begin{equation}
		P_1(v) \approx \frac{ \mathcal{N} \, c^{\frac{\beta}{\alpha}} \, \Delta^{\beta\left(\frac{1}{\alpha}-1 \right)} }{ \alpha \, v^{1+\frac{\beta}{\alpha}}} \exp\left[ \frac{\beta^2 \lambda^2 \sigma_t^2}{2\alpha^2} \right] \left( \frac{1 + r^{\frac{\beta}{\alpha}}}{2(1+r)^{\frac{\beta}{\alpha}}} \right)
	\label{eq:Next_Distribution}
	\end{equation}
	
	Repeating this procedure again to produce the distribution for the next generation, $P_2(v)$, we find that $P_2(v) \sim 1 / v^{1+\beta / \alpha^2}$. That is, evolving a distribution with a power law tail $1 / v^{1+\beta}$ results in a distribution with a power law tail $1 / v^{1+\beta / \alpha}$. Then by a simple inductive argument, after evolving our initial distribution $n$ times, we are left with a distribution whose tail is given by
	
	\begin{equation}
		P_0(v) \sim \frac{1}{v^{1+\beta}} \to \cdots \to P_n(v) \sim \frac{1}{v^{1+\beta / \alpha^n}} .
	\end{equation}
	
	\noindent This implies that, for $0 < \alpha < 1$, the power of the tail gets higher and the tail itself gets lighter. Thus for such policies, the stable distribution decays faster than any power law tail. On the other hand, for $\alpha > 1$, successive iterations only bring the tail of the distribution closer to $1 / v$. Though the distribution will never reach the non-normalizable distribution $1 / v$ in finite time, it follows that growth policies with supralinear growth asymptote to $1 / v$.
	
	When $\alpha = 1$, the power of the tail does not change from generation to generation. This suggests that the stable distribution should indeed have a power law tail, the only question is what the power of the tail should be. Now if we make the ansatz that the initial power-law distribution $P_0(v)$ is the actual tail of the stable distribution, then Eq. \ref{eq:Next_Distribution} requires that
	
	\begin{equation}
		\frac{1}{v^{1+\beta}} = \frac{ c^{\beta} }{ v^{1+\beta}} \left( \frac{1 + r^{\beta}}{2(1+r)^{\beta}} \right) \exp\left[ \frac{\beta^2 \lambda^2 \sigma_t^2}{2} \right],
	\end{equation}
	
	\noindent which is a constraint equation for $\beta$:
	
	\begin{equation}
		c = (1+r) \exp\left[-\frac{1}{2} \beta \lambda^2 \sigma_t^2 \right] \left(\frac{2}{ 1 + r^\beta } \right)^{\frac{1}{\beta}}.
	\label{eq:Power_Law_Exponent_Constraint}
	\end{equation}
	
	\noindent This equation for $\beta$ matches the constraint we found for the existence of the $k$th moment (Eq. \ref{eq:Moment_Constraint}), only now we have a constraint valid for real values of $\beta$ instead of just integer values, allowing us to plot the power law exponent as a function of the control coefficient $c$ and multiplicative noise amplitude $\lambda\sigma_t$ (Fig. \ref{fig:Power_Law_Exponent_Figure}). 
	
	\begin{figure}
		\centering
			\includegraphics[width=\linewidth]{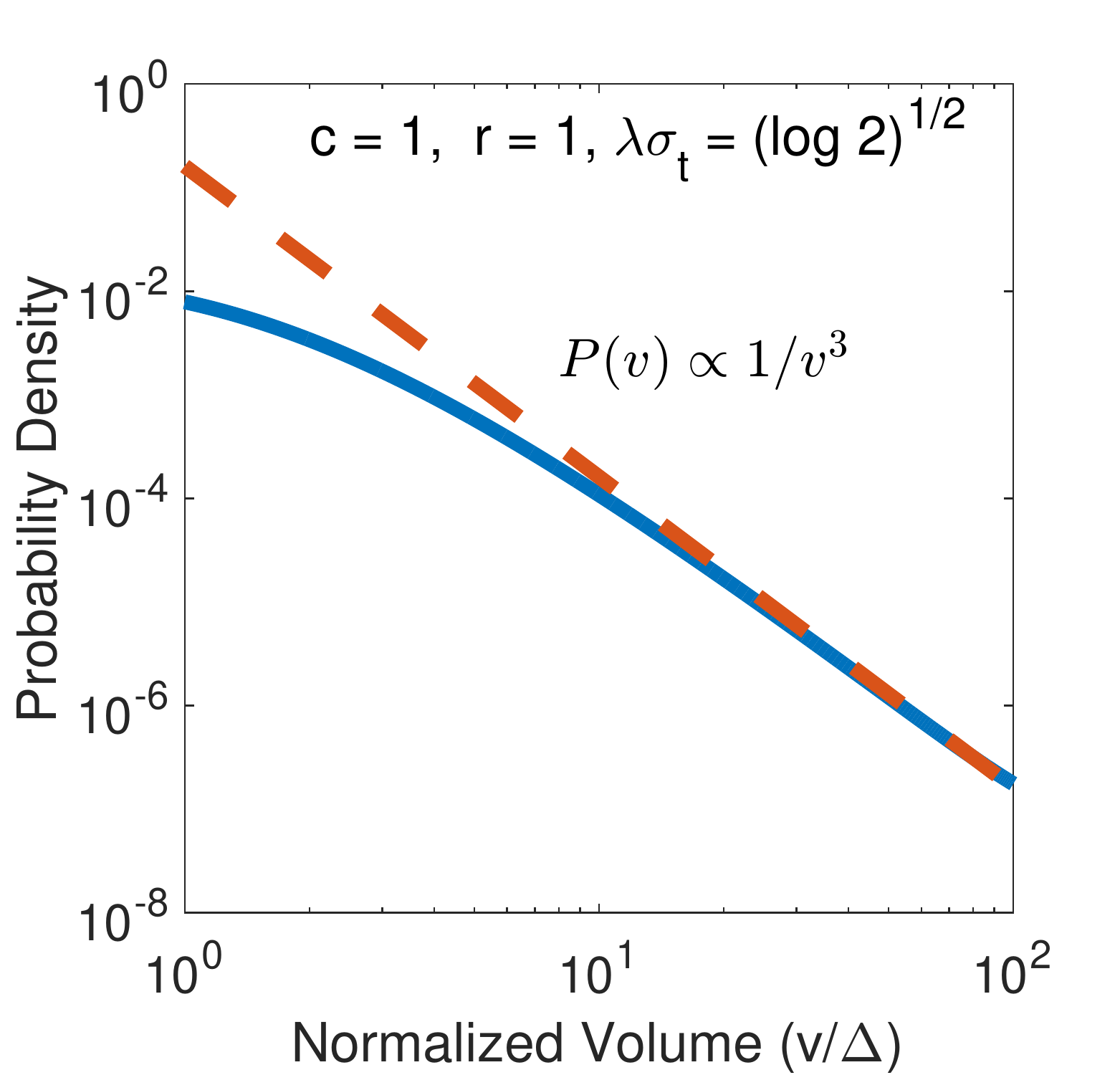}
		\caption{(Color online) The power-law tail for symmetric division ($r = 1$) under the policy $f(v) = \Delta + v$, i.e. $c = 1$, with multiplicative noise amplitude $\lambda\sigma_t = \sqrt{\log 2}$. As indicated in Eq. \ref{eq:No_Variance_Exists}, this is the smallest noise amplitude for which the variance ceases to exist. The solid blue line is the stable cell volume distribution (obtained via the discretization of the integral equation discussed in App. \ref{App:Discretization}) and the dashed red line represents the power law with exponent $1 + \beta = 3$ determined by Eq. \ref{eq:Power_Law_Exponent_Symmetric}.}
		\label{fig:Power_Law_Fit}
	\end{figure}
	
	\subsection{The Special Case of Symmetric Division}
	\label{Subsec:Special_Case}
	
		We can explicitly solve for $\beta$ in the case of symmetric division. Taking $r = 1$ in Eq. \ref{eq:Power_Law_Exponent_Constraint} yields a greatly simplified expression:
		
		\begin{equation}
			c = 2 \exp\left[-\frac{1}{2} \beta \lambda^2 \sigma_t^2 \right],
		\end{equation}
		
		\noindent which implies that $\beta$ is
		
		\begin{equation}
			\beta = \frac{2 \left( \log 2 - \log c \right)}{\lambda^2 \sigma_t^2}.
		\label{eq:Power_Law_Exponent_Symmetric}
		\end{equation}
		
		The power of the tail grows with both the multiplicative noise amplitude ($\lambda \sigma_t$) and the control coefficient $c$. Hence for a typical case with $c = 1$ and $\lambda \sigma_t = 0.2$, we have $\beta = 34.7$, which implies that the stable distribution is rather well behaved. But how much noise could a cell with $c = 1$ tolerate before having an undefined variance? In other words, how large can we take $\lambda \sigma_t$ while keeping $\beta \geq 2$? Solving Eq. \ref{eq:Power_Law_Exponent_Symmetric} for $\lambda \sigma_t$, we see that the bound on the noise is fairly generous,
		
		\begin{equation}
			\lambda \sigma_t \leq \sqrt{ \log 2 } \approx 0.8.
		\label{eq:No_Variance_Exists}
		\end{equation}
		
		\noindent We show in Fig. \ref{fig:Power_Law_Fit} that the tail of the stable distribution at this multiplicative noise level does indeed have $\beta = 2$.


\section{Stability Phase Diagram}
\label{Sec:Phase_Diagram}

	\begin{figure}[t!]
		\centering
			\includegraphics[width=\linewidth]{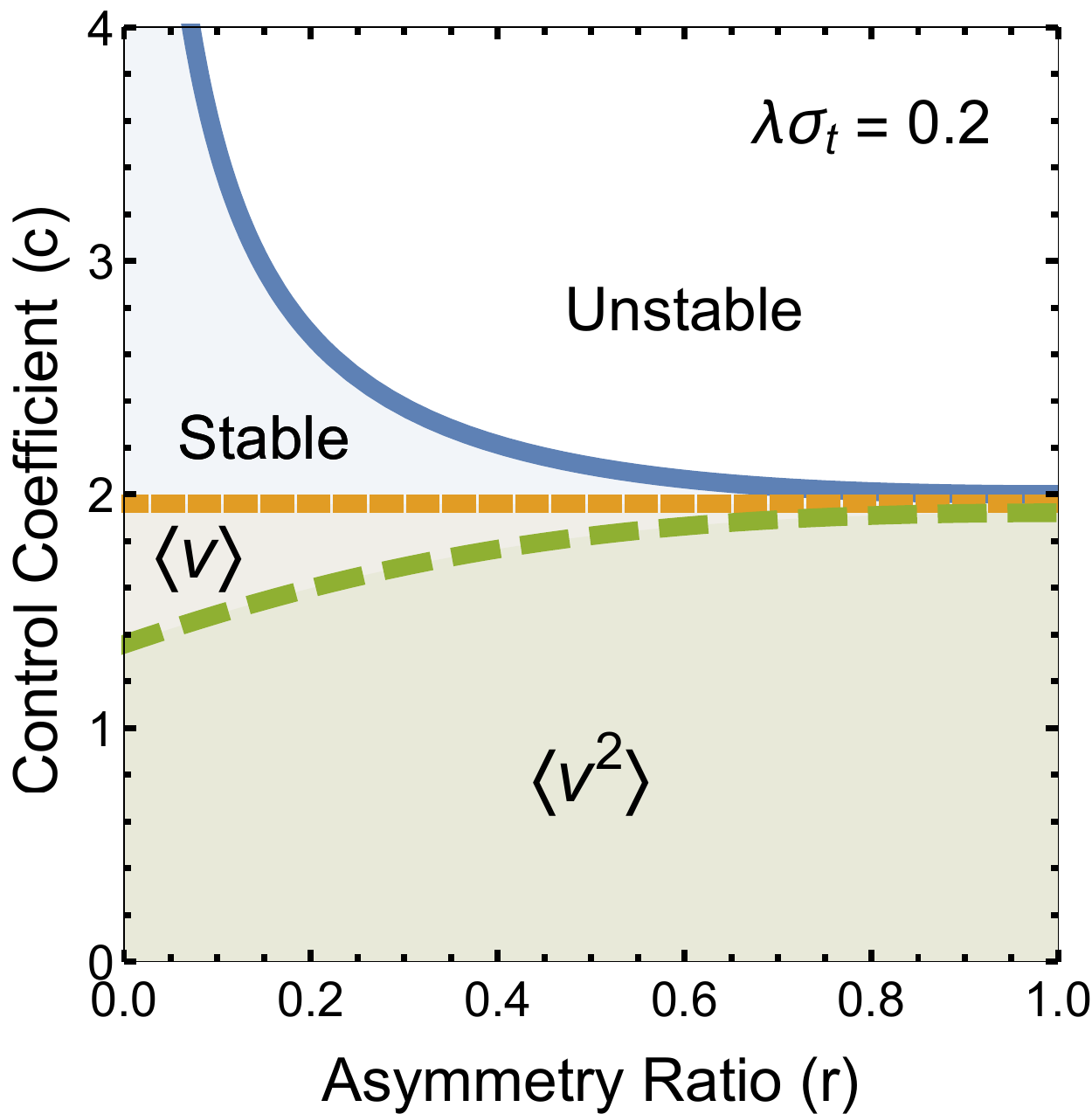}
		\caption{(Color online) Stability phase diagram for multiplicative noise with amplitude $\lambda \sigma_t = 0.2$. The labels $\langle v \rangle$ and $\langle v^2 \rangle$ represent the regions (defined by Eq. \ref{eq:Moment_Constraint}) in which the first and second moments exist, respectively. Note that the stable region includes the $\langle v \rangle$ region, which includes the $\langle v^2 \rangle$ region, and so on. The overall region of stability is described by the area under the curve given in Eq. \ref{eq:Stability_Bound} and is independent of the noise amplitude; however, the regions in which the various moments of the stable distribution exist do depend (weakly) on the noise. That increasing asymmetry (i.e. $r \to 0$) allows for larger values of the control coefficient $c$ is a consequence of the fact that, though the smaller cells produced in asymmetric division eventually grow large, they produce many more smaller cells as they grow. Thus the large number of small-volume cells compensated for the relatively small number of large cells.}
	\label{fig:Phase_Diagrams}
	\end{figure}

	The power-law constraint from Eq. \ref{eq:Power_Law_Exponent_Constraint} allows us to make a prediction about stability. In the limit as $\beta \to 0$, the tail of the stable cell volume distribution will edge closer and closer to $1 / v$, at which point the integral of the distribution will diverge logarithmically. Thus if the system is operating under conditions such that $\beta < 0$, then the cell volume distribution will not converge to a stable \emph{distribution} but instead asymptote to a non-normalizable function. We can then understand the stability of the system by taking the limit as $\beta \to 0$ in Eq. \ref{eq:Power_Law_Exponent_Constraint}, which implies that, for a fixed asymmetry ratio, the largest stable control coefficient ($\cmax^{(0)}$) is given by
	
	\begin{equation}
		\cmax^{(0)} = \frac{1+r}{\sqrt{r}}.
	\label{eq:Stability_Bound}
	\end{equation}
	
	\noindent Remarkably, this stability bound is only a function of the asymmetry ratio and \emph{not} dependent on the amplitude of the noise. Moreover, though it may not appear to be true at first glance, Eq. \ref{eq:Stability_Bound} is indeed invariant under the transformation $r \to 1/r$ (i.e. changing the arbitrary ``mother'' and ``daughter'' labels), as all our population-level results must be. We should note, however, that this is only true when $\Delta \neq 0$ (as we show in Sec. \ref{Sec:Proof}, there is no stable distribution when $\Delta = 0$). We combine this stability bound with our constraints for the moments (Eq. \ref{eq:Moment_Constraint}) to produce a stability phase diagram in Fig. \ref{fig:Phase_Diagrams} for multiplicative noise of amplitude $\lambda\sigma_t = 0.2$.
	
	It is rather interesting that increasing asymmetry ($r$ approaching 0) increases stability (i.e. the range of stable values of $c$). To gain some intuition as to why this occurs, we consider how a single cell (with $r$ very close to zero) grows into a stable population. Every generation, our original cell will get larger and larger, and so as time goes on this cell will grow to infinity. This is the mechanism by which the heavy tail arises. However, every generation it will spawn a very tiny cell (since $r$ is so small) that will also begin to grow larger and larger. But as these tiny cells grow, they spawn even tinier cells, and so at any given time, there exist many cells of small volume even though the older cells continue to swell in size.

	Now since $P(v)$ represents the volume distribution for this population, the relatively large number of small cells counteracts the effects of the relatively few large cells. The larger the asymmetry, the larger this disparity (and control coefficient $c$) can be while still maintaining a stable situation.


\section{Additive Models and Damage Control}
\label{Sec:Damage_Control}

	\begin{figure}
		\centering
			\includegraphics[width=\linewidth]{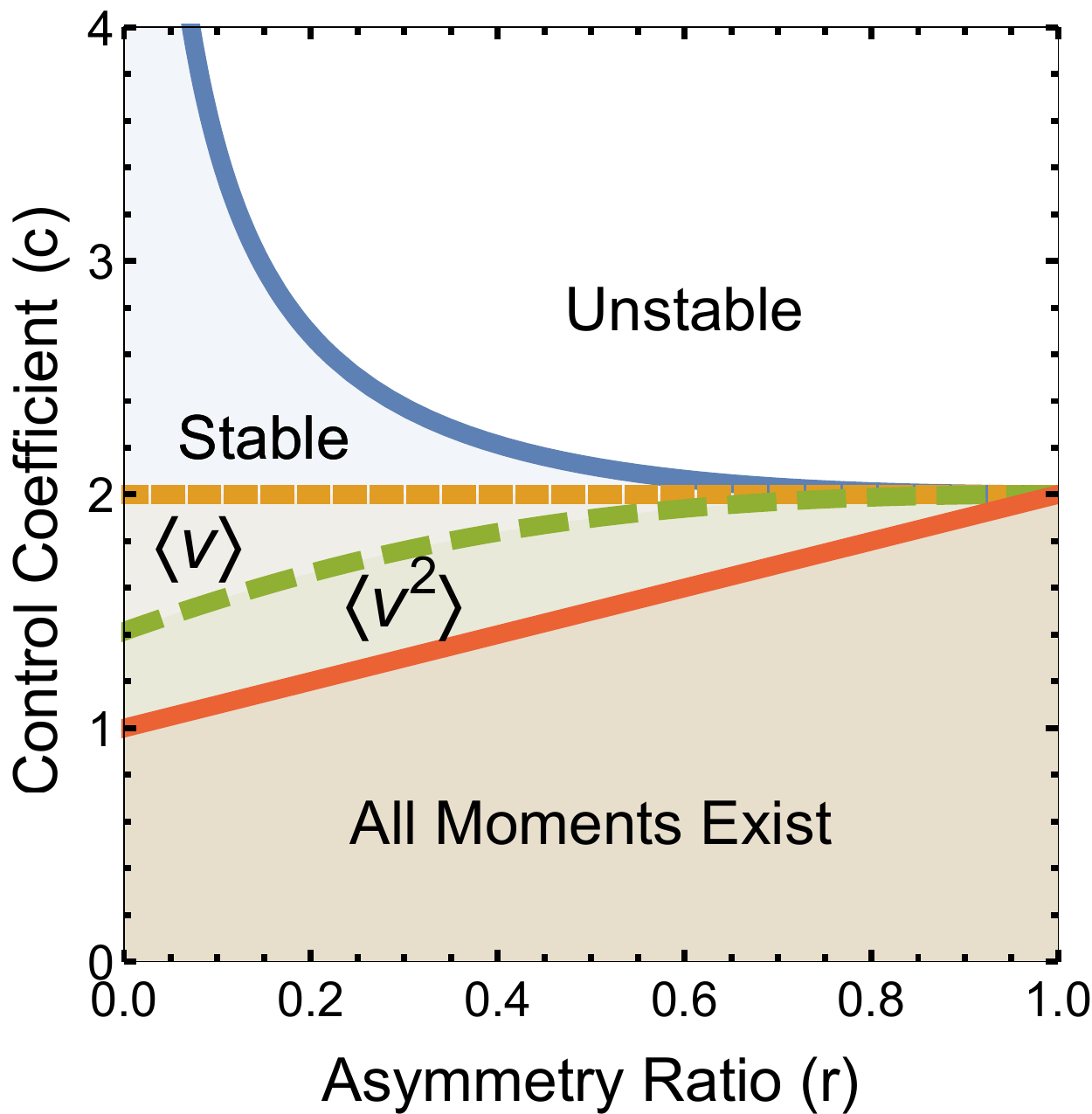}
		\caption{(Color online) Stability phase diagram for systems with no multiplicative noise and additive noise of any amplitude. Unlike the case of multiplicative noise (cf. Fig. \ref{fig:Phase_Diagrams}), the bounds on the moments do not depend on the noise amplitude, but only on the asymmetry ratio (Eq. \ref{eq:Additive_Bounds}). In addition, as we indicate in Eq. \ref{eq:Additive_All_Moments}, under additive models there is a region $c \leq 1 + r$ in which all moments of the stable distribution exist. }
	\label{fig:Phase_Diagrams_Additive}
	\end{figure}

	\begin{figure}
		\centering
			\includegraphics[width=\linewidth]{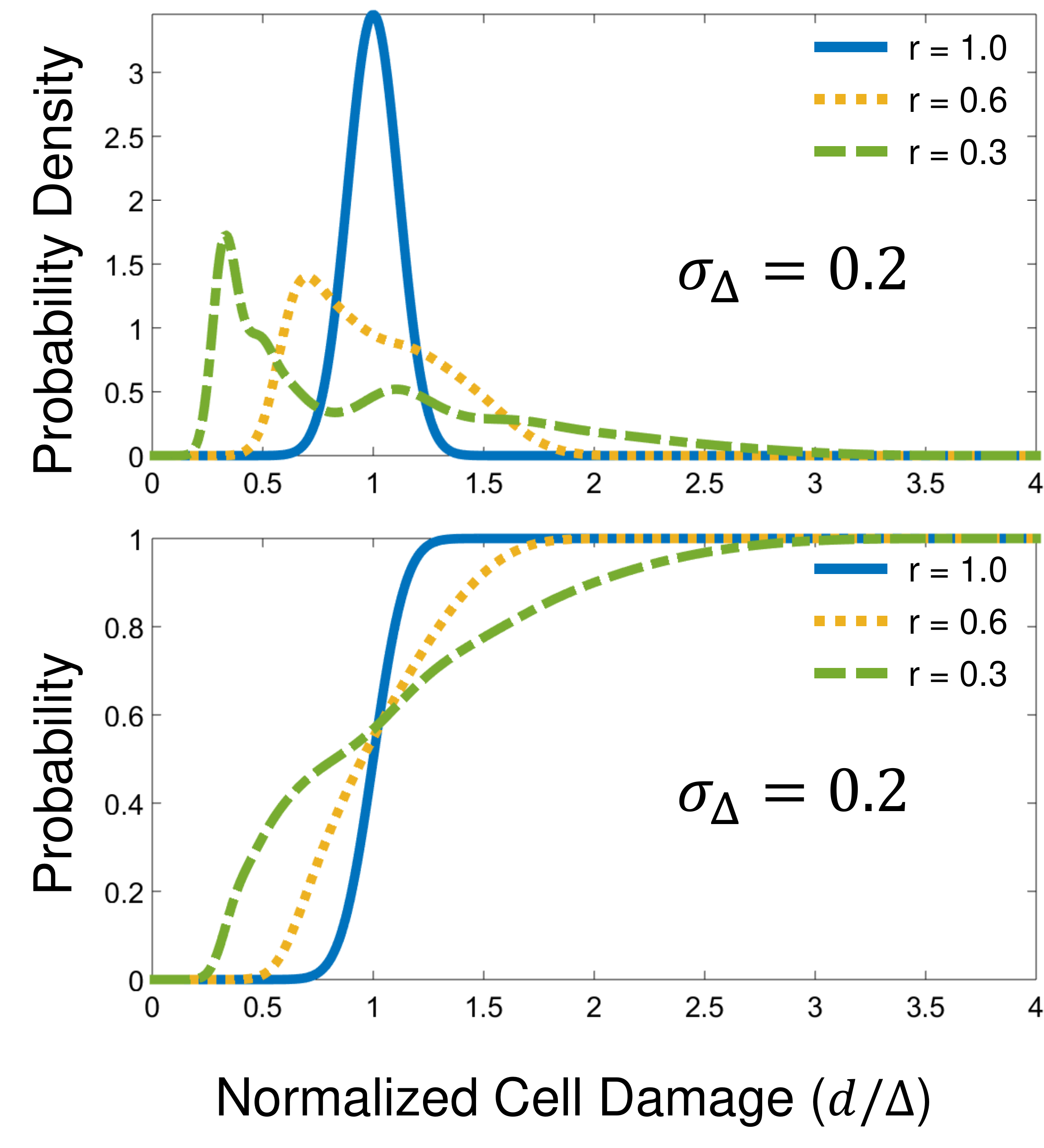}
		\caption{(Color online) Stable damage distributions (top) and the corresponding cumulative distributions (bottom) for various asymmetry ratios with additive noise amplitude $\sigma_\Delta$ = 0.2. In cases where the cells' damage tolerance threshold is below the mean damage incurred during growth ($\Delta$), the cells can increase the probability for their progeny to be under the threshold by dividing the damage asymmetrically. Though it may appear that the cumulative distributions meet at probability 1/2, they do not: the mean is the same for all values of $r$ (see Eq. \ref{eq:Damage_Mean}) but in general the median is not. We numerically solved for these distributions using a Fredholm Integral Equation Solver by K. Atkinson and L. Shampine \cite{atkinson2008algorithm}.}
	\label{fig:Stable_Damage_Distributions}
	\end{figure}
	
	In the case of additive models (i.e. models with no multiplicative noise), the results of Sec. \ref{Sec:Moments} still hold with all of the expectations over multiplicative noise set to unity, $\langle e^{k \lambda t^N} \rangle \to 1$. The stability constraint from Eq. \ref{eq:Stability_Bound} is also intact, though the behavior of the tail is quite different. Without multiplicative noise, the bounds on the control coefficient, $\cmax^{(k)}$,
	
	\begin{equation}
		c < (1+r) \, \left( \frac{2}{ 1 + r^k} \right)^{\frac{1}{k}} = \cmax^{(k)}.
	\label{eq:Additive_Bounds}
	\end{equation}
	
	\noindent do not tend to zero as $k \to \infty$. Instead we find that there exists a bound $\cmax^{(\infty)}$,
	
	\begin{equation}
		\cmax^{(\infty)} = \begin{cases} 1 + r &\mbox{ if } 0 \leq r \leq 1, \\ 1 + \frac{1}{r} &\mbox{ if } r \geq 1, \end{cases}
	\label{eq:Additive_All_Moments}
	\end{equation}
	
	\noindent such that if $c < \cmax^{(\infty)}$, then all moments of the stable distribution exist. Note that $\cmax^{(\infty)}$ is indeed still invariant under $r \to 1/r$. Furthermore, since $\cmax^{(\infty)}$ is always at least 1, when $c \leq 1$ it follows that all moments exist for any choice of $r$. Thus it appears that additive models with $c \leq 1$ are not heavy-tailed. We display this interesting behavior more visually in Fig. \ref{fig:Phase_Diagrams_Additive}. 
	
	It is also useful to pay special attention to the damage model (Eq. \ref{eq:Damage_Model}) in which the control coefficient $c$ is set to 1. In this case, the mean damage at birth $\langle d \rangle$ is simply the mean damage accrued per growth cycle,
	
	\begin{equation}
		\langle d \rangle = \Delta,
	\label{eq:Damage_Mean}
	\end{equation}
	
	\noindent and the variance of the damage simplifies to
	
	\begin{equation}
		\sigma_d^2 = \frac{\sigma_\Delta^2 (1+r^2) + 2 \Delta^2 (1-r)^2}{ 1 + r(4+r)},
	\label{eq:Damage_Variance}
	\end{equation}
	
	\noindent where $\Delta$ is the average amount of damage incurred per cycle and $\sigma_\Delta^2$ is the variance about this average. This agrees with Eq. 17 in the supplementary material of Ref. \cite{amir2014cell} for $r = 1$ (note the different noise convention). The variance in the damage at birth, $\sigma_d^2$, increases with increasing asymmetry, but in a very nontrivial way. As the asymmetry increases, the stable distribution first moves from unimodal to bimodal, and then as $r$ approaches 0, more and more peaks emerge (see Fig. \ref{fig:Stable_Damage_Distributions}). 
	
	There is, however, a potential benefit to this increasing spread. For example, say the cells can only tolerate an amount of damage $d^* < \Delta$. Then the fraction of the population with $d < d^*$ (which can be seen in the plots of the cumulative distribution in Fig. \ref{fig:Stable_Damage_Distributions}) is actually greater for greater asymmetry (smaller $r$). Naively, the best thing to do is to put all of the damage into one daughter cell, but this likely requires a complex mechanism. Thus an optimization problem arises: given some metric of the difficulty of effecting a certain amount of asymmetry, one can ask for the optimal asymmetry level $r^*$ that maximizes the fraction of the population with damage levels $d < d^*$. Of course we should note that none of our modeling takes into account cell death or the effects of damage on cell growth, and so this argument is merely heuristic.
	\newpage


\section{Conclusion}

	By treating the abstract mathematical problem of stochastic division, we are able to eschew the delicate and complicated matter of the actual biological implementation while broadening the scope of our results. From an experimental perspective, the value of our analysis rests in our explicit formulae for the readily-measurable moments of the stable distribution and the corresponding stability phase diagram. Our analysis also provides intriguing characterizations of the existence and uniqueness of the stable distribution and its power-law tail.



\begin{acknowledgments}
	We would like to thank Andrej Ko\v{s}mrlj and Naama Brenner for their helpful comments. This research was conducted with Government support under and awarded by DoD, Air Force Office of Scientific Research, National Defense Science and Engineering Graduate (NDSEG) Fellowship, 32 CFR 168a.
\end{acknowledgments}
\clearpage


\appendix

\section{Equivalence of the generation-based and time-based stable cell birth volume distributions}
\label{App:Generations_vs_Time}

	Though it is analytically convenient to work with a generation-based analysis, in practice it is useful to consider the distribution for the birth size of cells present at a given time. Here we show that, in the long-time limit, the stable birth volume distribution for a given generation of cells is also the stable birth volume distribution for cells present at time $t$. To see this, we write our distribution for the birth volume of a cell present at time $t$, $P_t(v)$, in terms of the birth volume distribution for a given generation, $P_n(v)$,

	\begin{equation}
		P_t(v) = \sum_{n=0}^\infty P_n(v) \, P(n | t).
	\label{eq:Marginalize_Generations}
	\end{equation}
	
	\noindent where $P(n | t)$ is probability for the cell to belong to the $n$th generation at time $t$. For long times $t \gg \lambda^{-1}$, it will be very unlikely that our chosen cell is from an early generation, so $P(n|t)$ will be concentrated around high generation numbers. Then since we know that $P_n(v) \to P(v)$ for large $n$, it follows that we can approximate the sum by
	
	\begin{equation}
		P_t(v) \approx P(v) \sum_{n=0}^\infty P(n | t) = P(v),
	\end{equation}
	
	\noindent which, in the limit as $t \to \infty$, should give us the desired correspondence between the stable birth volume distribution for a given generation and at a given time,
	
	\begin{equation}
		\lim_{t\to\infty} P_t(v) = P(v).
	\label{eq:Stable_Equivalence}
	\end{equation}


\section{Discretization and consistency}
\label{App:Discretization}

	\begin{figure}[t!]
		\centering
			\includegraphics[width=\linewidth]{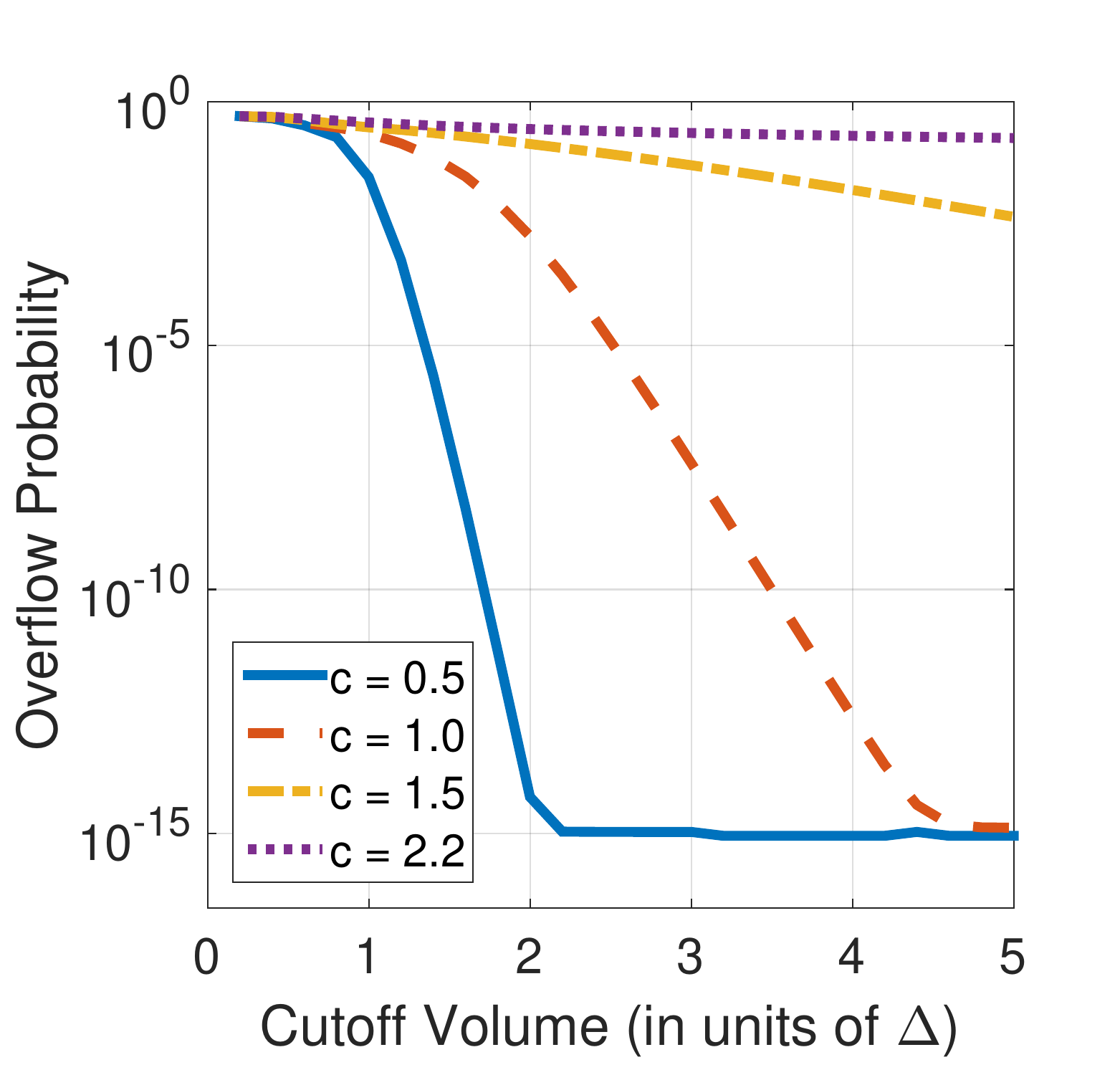}
			\caption{(Color online) Probability for a cell to be in the overflow bin as a function of cutoff volume for an affine linear policy $f(v) = \Delta + c \, v$ with asymmetry ratio $r = 0.6$ under the discretization choice (Eq. \ref{eq:Discretization_Choice}). All volumes are measured in units of $\Delta$ and the volume bins have a width of $\Delta \times 10^{-3}$. Note that the y-axis is logarithmic. The probabilities correspond to the final element of the eigenvector of $K$ with eigenvalue 1. For $c = 0.5, 1.0$ and $1.5$, the overflow probability falls off exponentially with the cutoff volume $\mathcal{V} / \Delta$ until the overflow probability becomes small enough to be sensitive to numerical errors. The policy with $c = 2.2$ operates in an unstable regime (see Fig. \ref{fig:Phase_Diagrams}) and the overflow probability falls off extremely slowly with the cutoff, reflecting the non-normalizability of the unstable distribution. This suggests that our choice in discretizing Eq. \ref{eq:Eigenvalue_Equation} is consistent.}
	\label{fig:Overflow_Probability_Figure}
	\end{figure}
	
	Here we discuss the process of discretizing Eq. \ref{eq:Eigenvalue_Equation} to produce the finite-dimensional matrix equation of Eq. \ref{eq:Discrete_Eigenvalue_Equation}. Instead of considering the probability density $P_n(v)$ on the half-line, we define a volume cutoff $\mathcal{V}$ and divide the interval $[0,\mathcal{V}]$ into a set of $N$ volume bins, allowing us to define a probability vector $\pvec_n$ whose first $N$ elements represent the probability that a cell has a volume in a given bin during generation $n$ and whose last element represents the probability for a cell in generation $n$ to have a volume $v \geq \mathcal{V}$. We can construct the elements of the matrix $K$ (excluding the final row and final column) by taking integrals over the kernel (Eq. \ref{eq:Kernel}),
	
	\begin{equation}
		K_{ij} = \frac{N}{\mathcal{V}} \int_{(i-1) \frac{\mathcal{V}}{N}}^{i \frac{\mathcal{V}}{N}} dv \int_{(j-1) \frac{\mathcal{V}}{N}}^{j \frac{\mathcal{V}}{N}} dv' \; K(v, v').
	\label{eq:Discrete_Kernel}
	\end{equation}
	
	\noindent Owing to the different range of integration for the overflow bin, the matrix elements for the final row are slightly different, but still based on our continuous kernel,
	
	\begin{equation}
		K_{(N+1), \, j} = \frac{N}{\mathcal{V}} \int_{\mathcal{V}}^\infty dv \int_{(j-1) \frac{\mathcal{V}}{N}}^{j \frac{\mathcal{V}}{N}} dv' \; K(v,v').
	\end{equation}
	
	\noindent This ensures that the first $N$ columns of $K$ sum to 1, which is necessary for $K$ to be a stochastic matrix.
	
	As for the final column, $K_{i,(N+1)}$ is the probability for a cell in the overflow bin to end up in the $i$th bin during the next generation. However, simply knowing that a cell is in the overflow bin does not give much information about its actual volume: we no longer have the option of assuming the cell is equally likely to be anywhere in the bin since the bin $[\mathcal{V}, \infty)$ has infinite size. Thus it appears that there is some ambiguity in discretizing Eq. \ref{eq:Eigenvalue_Equation}. 
	
	Fortunately, the ambiguity in choosing $K_{i,(N+1)}$ is irrelevant when Eq. \ref{eq:Eigenvalue_Equation} admits a stable distribution. As we note in Sec. \ref{Sec:Proof}, the existence of a stable distribution implies that we can always make the probability for being in the overflow bin arbitrarily small by choosing a large enough cutoff $\mathcal{V}$. Since the $K_{i,(N+1)}$ will always be multiplied by this infinitesimal probability, our choice should not matter so long as we ensure that $K$ is still a non-negative, connected stochastic matrix so that the Perron-Frobenius theorem is still applicable. 
	
	On the other hand, if the system is unstable, the overflow probability will not decay to zero for any choice of $K_{i,(N+1)}$. If it were to decay, then we could construct a stable distribution by smoothing the discrete solution, in direct contradiction to our assumption of stability. Thus a system is stable if and only if the overflow probability in its discrete analogue decays to zero with the cutoff. This is true for any choice of $K_{i,(N+1)}$ that preserves the non-negativity, connectedness and stochasticity of $K$, though the choice may affect the speed and accuracy with which Eq. \ref{eq:Discrete_Eigenvalue_Equation} can be solved.
	
	Fig. \ref{fig:Overflow_Probability_Figure} illustrates this method for a particularly simple choice of $K_{i,(N+1)}$,
	
	\begin{equation}
		K_{i,(N+1)} = \frac{1}{N+1}.
	\label{eq:Discretization_Choice}
	\end{equation}
	
\phantom{heehaw}
\newpage
\phantom{hmm}

\section{Calculation of the $k$th moment of the stable volume distribution}
\label{App:kth_Moment}
	
	Here we calculate the $k$th moment of the stable cell volume distribution under the affine linear policy of Eq. \ref{eq:Affine_Linear_Policy}. Under such a policy, Eq. \ref{eq:Kth_Moment} becomes
	
	\begin{equation}
		\left\langle v^k \right\rangle = \frac{1 + r^k}{2(1+r)^k} \left\langle \left( (\Delta + c \, v) \, e^{\lambda t^N} + v^N \right)^k \right\rangle.
	\end{equation}
	
	\noindent We then use the Binomial theorem to expand the term inside the expectation operator,
	
	\begin{multline}
		\left( (\Delta + c \, v) \, e^{\lambda t^N} + v^N \right)^k \\ = \sum_{j=0}^k \binom{k}{j} (v^N)^{k-j} \, e^{j \lambda t^N} \, (\Delta + c \, v)^j,
	\end{multline}
	
	\noindent and then similarly for the factor of $(\Delta + c \, v)^j$,
	
	\begin{equation}
		(\Delta + c \, v)^j = \Delta^j \sum_{l=0}^j \binom{j}{l} \left(\frac{c \, v}{\Delta}\right)^l,
	\end{equation}
	
	\noindent which then results in the following expansion for $\langle v^k \rangle$,
	
	\begin{widetext}
	\begin{equation}
		\left\langle v^k \right\rangle = \frac{\Delta^k (1 + r^k)}{2(1+r)^k} \sum_{j=0}^k \sum_{l=0}^j \binom{k}{j} \binom{j}{l} \left\langle \left( \frac{v^N}{\Delta} \right)^{k-j} \right\rangle \left\langle e^{j \lambda t^N} \right\rangle \left\langle \left( \frac{c \, v}{\Delta} \right)^l \right\rangle .
	\label{eq:Recursive_Moment_Equation}
	\end{equation}
	
		\noindent To solve this expression for $\langle v^k \rangle$, we have to move all the $\langle v^k \rangle$ terms to the left hand side. First we re-index the sum,
		
		\begin{equation}
			\left\langle v^k \right\rangle = \frac{\Delta^k (1 + r^k)}{2(1+r)^k} \sum_{l=0}^k \left\langle \left( \frac{c \, v}{\Delta} \right)^l \right\rangle \sum_{j=l}^k \binom{k}{j} \binom{j}{l} \left\langle \left( \frac{v^N}{\Delta} \right)^{k-j} \right\rangle \left\langle e^{j \lambda t^N} \right\rangle ,
		\label{eq:Reindexed_Moment_Equation}
		\end{equation}
		
		\noindent so that we can now extract the term depending on the $k$th moment,
		
		\begin{equation}
			\left\langle v^k \right\rangle = \frac{\Delta^k (1 + r^k)}{2(1+r)^k} \left( \frac{c^k}{\Delta^k} \left\langle e^{k \lambda t^N} \right\rangle  \left\langle v^k \right\rangle +  \sum_{l=0}^{k-1} \left\langle \left( \frac{c \, v}{\Delta} \right)^l \right\rangle \sum_{j=l}^k \binom{k}{j} \binom{j}{l} \left\langle \left( \frac{v^N}{\Delta} \right)^{k-j} \right\rangle \left\langle e^{j \lambda t^N} \right\rangle  \right) .
		\end{equation}
		
		\noindent Then we can explicitly solve for $\langle v^k \rangle$,
		
		\begin{equation}
			\left\langle v^k \right\rangle = \frac{\Delta^k (1+r^k)}{2(1+r)^k-\left( 1 + r^k \right) c^k \left\langle e^{k \lambda t^N} \right\rangle } \sum_{l=0}^{k-1} \left\langle \left( \frac{c \, v}{\Delta} \right)^l \right\rangle \sum_{j=l}^k \binom{k}{j} \binom{j}{l} \left\langle \left( \frac{v^N}{\Delta} \right)^{k-j} \right\rangle \left\langle e^{j \lambda t^N} \right\rangle .
		\label{eq:Explicit_Kth_Moment}
		\end{equation}
		
		\noindent Starting from the explicit formula for the mean (Eq. \ref{eq:Stable_Mean}), this equation is sufficient to recursively generate as many moments of the stable distribution as we wish. 
		 
		\end{widetext}
		\phantom{why}
		\clearpage

\section{Simulating asymmetric cell division}

	We performed a Monte Carlo simulation of asymmetric cell division to compare with our analytical results from Sec. \ref{Sec:Moments}. These simulations consisted of repeated rounds of stochastic growth and division with fixed asymmetry ratio $r$. We also prevented cells from shrinking during the growing phase (which is possible, though rare, due to noise). Table \ref{Table:Moment_Comparison} shows the close numerical agreement between the simulations and our predictions for the mean and variance from Eqs. \ref{eq:Stable_Mean} and \ref{eq:Stable_Variance}.

		\begin{table}[t!]
		\caption{Comparison of predicted and simulated moments of the stable cell volume distribution for affine linear growth policies (Eq. \ref{eq:Affine_Linear_Policy}) under a range of asymmetry ratios ($r$), control coefficients ($c$) and multiplicative noise strengths ($\lambda\sigma_t$). The values reported for the normalized mean $\langle v / \Delta \rangle$ and variance $\sigma_v^2 / \Delta^2$ are the theoretical values obtained from Eqs. \ref{eq:Stable_Mean} and \ref{eq:Stable_Variance}. We compare each of these values to cell growth simulations ($N = 10^4$) in which the initial set of cells were allowed to grow for 10 generations, after which 500 cells would be randomly chosen to serve as seeds for another cycle of growth. We repeated this cycling procedure 9 times to make for 10 full cycles of growth. We also prevented cells from shrinking by forcing cells that would have shrunk to not change volume at all. Nonetheless, the relative error in both the mean ($\sim 10^{-4}$) and variance ($\sim 10^{-3}$) is small. }
		
		\begin{tabular}{|c|c|c|c|c|c|c|}
		\hline
		\multicolumn{1}{|c|}{$r$} & \multicolumn{1}{c|}{$c$} & \multicolumn{1}{c|}{$\lambda \sigma_t$} & \multicolumn{1}{c|}{$\left\langle v \right\rangle / \Delta$} & \multicolumn{1}{c|}{$\textrm{RelErr}\left[ \left\langle v \right\rangle \right]$} & \multicolumn{1}{c|}{$ \sigma_v^2 / \Delta^2 $} & \multicolumn{1}{c|}{$\textrm{RelErr}\left[ \sigma_v^2 \right]$} \\ \hline
		0.6 & 0.5 & 0.1 & 0.67 & $2.12 \times 10^{-4}$ & 0.035 & $7.51 \times 10^{-4}$ \\ \hline
		0.6 & 0.5 & 0.2 & 0.68 & $8.88 \times 10^{-4}$ & 0.053 & $1.94 \times 10^{-3}$ \\ \hline
		0.6 & 1.0 & 0.1 & 1.01 & $2.83 \times 10^{-4}$ & 0.102 & $1.17 \times 10^{-3}$ \\ \hline
		0.6 & 1.0 & 0.2 & 1.04 & $6.62 \times 10^{-4}$ & 0.161 & $2.60 \times 10^{-3}$ \\ \hline
		0.6 & 1.5 & 0.1 & 2.04 & $1.31 \times 10^{-3}$ & 0.781 & $5.67 \times 10^{-3}$ \\ \hline
		0.6 & 1.5 & 0.2 & 2.17 & $2.54 \times 10^{-3}$ & 1.417 & $1.29 \times 10^{-2}$ \\ \hline
		1.0 & 0.5 & 0.1 & 0.67 & $2.13 \times 10^{-4}$ & 0.0048 & $2.93 \times 10^{-3}$  \\ \hline
		1.0 & 0.5 & 0.2 & 0.68 & $4.73 \times 10^{-4}$ & 0.0205 & $3.80 \times 10^{-3}$ \\ \hline
		1.0 & 1.0 & 0.1 & 1.01 & $2.81 \times 10^{-4}$ & 0.0138 & $3.12 \times 10^{-3}$ \\ \hline
		1.0 & 1.0 & 0.2 & 1.04 & $5.87 \times 10^{-4}$ & 0.0607 & $3.84 \times 10^{-3}$ \\ \hline
		1.0 & 1.5 & 0.1 & 2.04 & $9.62 \times 10^{-4}$ & 0.0982 & $5.34 \times 10^{-3}$ \\ \hline
		1.0 & 1.5 & 0.2 & 2.17 & $2.23 \times 10^{-3}$ & 0.493 & $1.03 \times 10^{-2}$ \\ \hline
		\end{tabular}
	\label{Table:Moment_Comparison}
	\end{table}
	
	\phantom{that's the stuff}
	\newpage

\section{The effects of stochasticity in the asymmetry ratio}
\label{App:Stochastic_Division}

	\begin{figure}[t!]
		\centering
			\includegraphics[width=\linewidth,trim={0 13cm 0 0},clip]{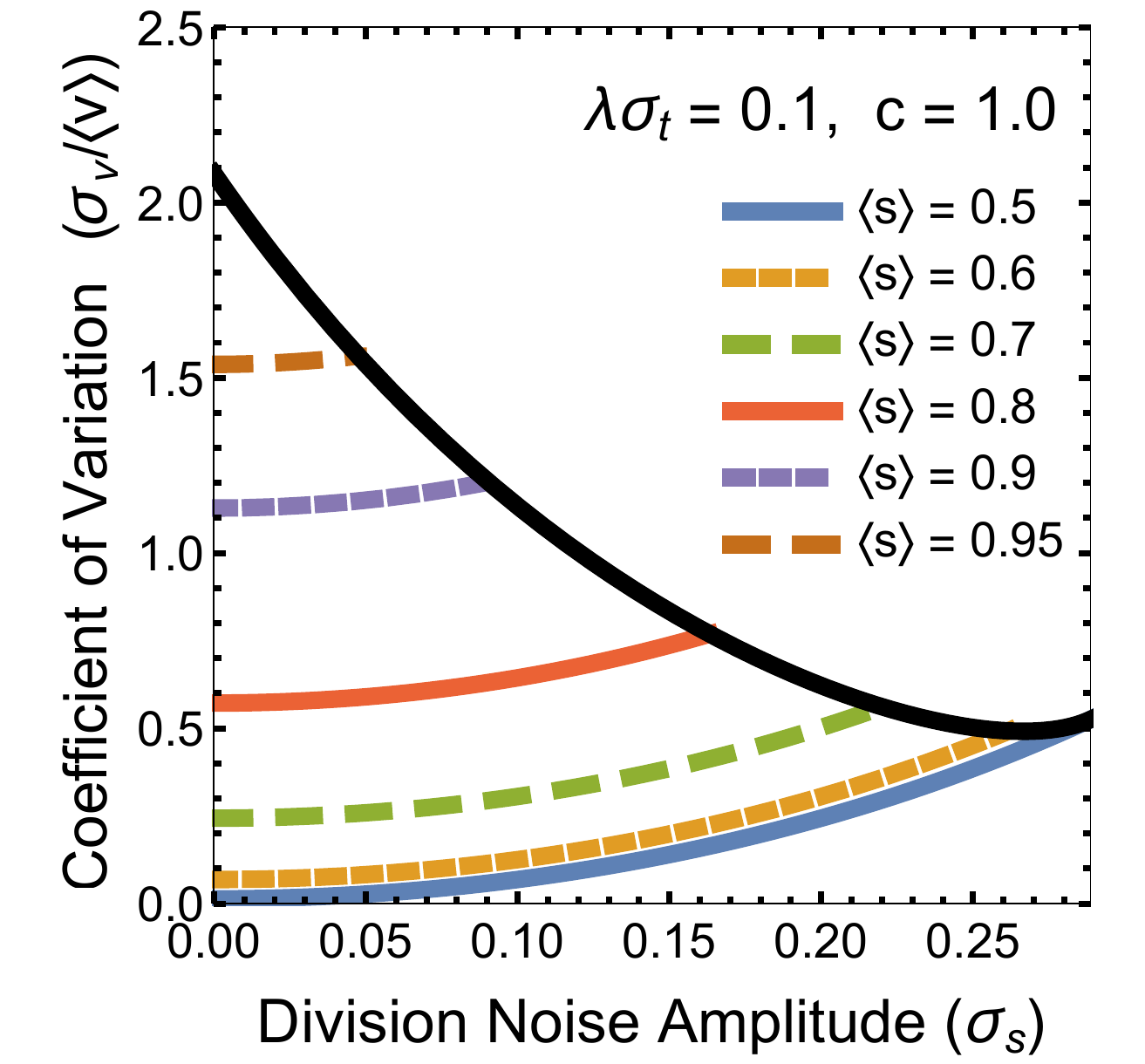}
		\caption{(Color online) The effects of a stochastic division fraction on the coefficient of variation (CV) of the stable cell volume distribution under the policy $f(v) = \Delta + c \, v$. Note that for a fixed mean division fraction $\langle s \rangle$, the corresponding variance $\sigma_s$ must be bounded above as shown in Eq. \ref{eq:Variance_Bounds}. The black line indicates the value of the CV for the maximum possible $\sigma_s$ for given $\langle s \rangle$. Note that the CV at $\sigma_s = 0$ corresponds to the CV for a system with fixed division fraction (or asymmetry ratio).} 
	\label{fig:Variance_Under_Stochastic_Division}
	\end{figure}
	
	If we allow for a stochastic asymmetry ratio, the primary difference in our development in Sec. \ref{Sec:Moments} is that Eq. \ref{eq:Kth_Moment} now involves expectations of functions of $r$,

	\begin{multline}
		\langle v^k \rangle =  \frac{1}{2} \left( \left\langle \frac{r^k}{(1+r)^k} \right\rangle + \left\langle \frac{1}{(1+r)^k} \right\rangle \right) \\ \times \left\langle \left( f(v) \, e^{\lambda t^N} + v^N \right)^k \right\rangle.
	\label{eq:Kth_Moment_Stochastic}
	\end{multline}
	
	\noindent However, it is generally inconvenient to compute moments with $1+r$ in the denominator. Instead, we first define the division fraction $s = 1 / (1+r)$ such that

	\begin{equation}
		v^M = s \, v^G, \qquad v^D = (1-s) \, v^G,
	\label{eq:Division_Factor}
	\end{equation}
	
	\noindent which allows us to write Eq. \ref{eq:Kth_Moment_Stochastic} as

	\begin{multline}
		\langle v^k \rangle =  \frac{1}{2} \left( \left\langle s^k \right\rangle + \left\langle (1-s)^k \right\rangle \right) \\ \times \left\langle \left( f(v) \, e^{\lambda t^N} + v^N \right)^k \right\rangle.
	\end{multline}
	
	\noindent For affine linear policies, we can use this to explicitly obtain the mean and variance. Since $\langle s \rangle + \langle 1 - s \rangle = 1$, the expression for the mean is the same as in Eq. \ref{eq:Stable_Mean}; however, the expression for the variance does change,
	
	\phantom{opera}
	\newpage
	
	\begin{widetext}
	\begin{equation}
		\sigma_v^2 = \frac{\Delta^2 \langle e^{2\lambda t^N} \rangle}{ \frac{2}{ \langle s^2 \rangle + \langle (1-s)^2 \rangle} - c^2 \langle e^{2\lambda t^N} \rangle} \left( \frac{\left\langle (v^N)^2 \right\rangle}{\Delta^2 \langle e^{2\lambda t^N} \rangle} + \frac{4 \langle e^{2\lambda t^N} \rangle - \frac{2 \langle e^{\lambda t^N} \rangle^2}{\langle s^2 \rangle + \langle (1-s)^2 \rangle}}{(2 - c \langle e^{\lambda t^N} \rangle)^2 \, \langle e^{2\lambda t^N} \rangle} \right).
	\label{eq:Stable_Variance_2}
	\end{equation}
	
	\noindent We can also provide an expression for the $k$th moment (analogous to Eq. \ref{eq:Explicit_Kth_Moment}),
	
	\begin{equation}
			\left\langle v^k \right\rangle = \frac{\Delta^k}{\frac{2}{\langle s^k \rangle + \langle (1-s)^k \rangle} - c^k \left\langle e^{k \lambda t^N} \right\rangle } \sum_{l=0}^{k-1} \left\langle \left( \frac{c \, v}{\Delta} \right)^l \right\rangle \sum_{j=l}^k \binom{k}{j} \binom{j}{l} \left\langle \left( \frac{v^N}{\Delta} \right)^{k-j} \right\rangle \left\langle e^{j \lambda t^N} \right\rangle ,
	\label{eq:Explicit_Kth_Moment_2}
	\end{equation}
	\end{widetext}
	
	\noindent from which we derive a necessary constraint on the control coefficient for the existence of the $k$th moment (i.e. the analogue of Eq. \ref{eq:Moment_Constraint}),
	
	\begin{equation}
			c < \left( \frac{2}{\left( \langle s^k \rangle + \langle (1 - s)^k \rangle \right) \left\langle e^{k \lambda t^N} \right\rangle } \right)^{\frac{1}{k}}.
	\label{eq:Growth_Constraint_Stochastic}
	\end{equation}

	Now in order to proceed, we must specify the distribution for $s$. Given that $s \in [0,1]$, it is not quite appropriate to use a Gaussian. Thus we instead assume a beta distribution,

	\begin{equation}
		P(s) = \frac{s^{\alpha-1}(1-s)^{\beta-1}}{\textrm{B}[\alpha,\beta]},
	\label{eq:Beta_Distribution}
	\end{equation}
	
	\noindent where the parameters $\alpha, \beta \geq 1$ are related to the mean $\bar{s}$ and variance $\sigma_s^2$ by
	
	\begin{equation} \begin{split}
		\alpha &= \bar{s} \left( \frac{\bar{s}(1-\bar{s})}{\sigma_s^2} - 1 \right), \\
		\beta &= (1-\bar{s}) \left( \frac{\bar{s}(1-\bar{s})}{\sigma_s^2} - 1 \right).
	\end{split} \end{equation}
	
	\noindent Note that we are not free to choose the mean and variance arbitrarily. If we fix the mean $\bar{s} \in [0,1]$, then the standard deviation is bounded below by 0 and above by
	
	\begin{equation}
		\sigma_s \leq \begin{cases} \bar{s} \sqrt{\frac{1-\bar{s}}{1+\bar{s}}} &\mbox{ for } 0 \leq \bar{s} \leq \frac{1}{2}, \\ (1-\bar{s}) \sqrt{\frac{\bar{s}}{2-\bar{s}}} &\mbox{ for } \frac{1}{2} \leq \bar{s} \leq 1. \end{cases}
	\label{eq:Variance_Bounds}
	\end{equation}
	
	\noindent Nevertheless, by taking $\alpha$ and $\beta$ large with $\bar{s}$ fixed, the beta distribution will begin to resemble a highly-peaked Gaussian. Moreover, for our purposes the beta distribution is convenient as it allows us to explicitly calculate the relevant moments,
	
	\begin{equation}
		\langle s^k \rangle = \frac{\textrm{B}[\alpha + k, \beta]}{\textrm{B}[\alpha, \beta]}, \quad \langle (1-s)^k \rangle = \frac{\textrm{B}[\alpha, \beta + k]}{\textrm{B}[\alpha, \beta]}.
	\end{equation}
	
	\noindent We use these expressions in Eq. \ref{eq:Stable_Variance_2} to plot (in Fig. \ref{fig:Variance_Under_Stochastic_Division}) the coefficient of variation ($\textrm{CV} = \sigma_v / \langle v \rangle$) for the stable volume distribution for different average division fractions, $\langle s \rangle = \bar{s}$, and variances, $\sigma_s^2$.

	We can also use these explicit moments to write our constraint on the control coefficient (Eq. \ref{eq:Growth_Constraint_Stochastic}) as
	
	\begin{equation}
			c < \left( \frac{2 \, \textrm{B}[\alpha, \beta]}{\left( \textrm{B}[\alpha + k,\beta] + \textrm{B}[\alpha,\beta + k] \rangle \right) \left\langle e^{k \lambda t^N} \right\rangle } \right)^{\frac{1}{k}}.
	\label{eq:Moment_Constraint_Stochastic}
	\end{equation}
	
	\noindent Assuming the above expression holds for arbitrary $k$, we can take the limit as $k \to 0$ to obtain the stability curve just as we did in Eq. \ref{eq:Stability_Bound}. However, in this case the stability curve is significantly more complicated,
	
	\begin{equation}
		c^{(0)}_\textrm{max} = \exp\left[ \psi(\alpha + \beta) - \frac{1}{2} \left( \psi(\alpha) + \psi(\beta) \right) \right],
	\end{equation}
	
	\noindent where $\psi(\cdot)$ is the digamma function. In general this stochasticity tightens the control coefficient constraints; however, the effect is negligible for the lower moments.

\section{Investigating a mother/daughter-dependent model}
\label{App:Mother_Daughter_Model}

	In the main text we assume that the distinction between mothers and daughters is only a matter of which cell is born larger; however, the differences may run deeper. It is believed \cite{turner2012cell, colman2001yeast, weiss2002saccharomyces, cosma2004daughter, di2009daughter, laabs2003ace2} that newly-born daughter cells employ a different growth policy than do experienced mother cells. Here, we account for this by allowing for two different growth policies, $f^M(v)$ and $f^D(v)$, for mother and daughter cells, respectively. Now if we know the birth size distributions for the mother and daughter cells for generation $n$, $P_n^M(v)$ and $P_n^D(v)$, what are our birth size distributions for mothers and daughters in the next generation, $P_{n+1}^M(v)$ and $P_{n+1}^D(v)$?
	
	If we pick one cell from the population of the $(n+1)$th generation, it has four possible lineages. If its mother was experienced (i.e., not a first-time mother), then it could either be the resulting mother cell ($M \to M$) or the resulting daughter cell ($M \to D$). On the other hand, if its mother was a first-time mother, then it can either be the now-experienced mother cell ($D \to M$) or the daughter cell ($D \to D$). We can then write the distributions for mother and daughter cell size as
	
	\begin{widetext}
	\begin{equation} \begin{split}
		P_{n+1}^M\left( v \right) &= P \left( v_{n+1}^M = v \, \middle| M \to M \right) P\left( M \to M \right) + P\left( v_{n+1}^M = v \, \middle| D \to M \right) P\left( D \to M \right), \\
		P_{n+1}^D\left( v \right) &= P\left( v_{n+1}^D = v \, \middle| M \to D \, \right) P\left( M \to D \, \right) + P\left( v_{n+1}^D = v \, \middle| D \to D \right) P\left( D \, \to D \,\right).
	\end{split} \end{equation}
	
	\noindent Since at any given time there are an equal number of mothers and daughters, we have

	\begin{equation} \begin{split}
		P_{n+1}^M\left( v \right) &= \frac{1}{2} \left[ P\left( v_{n+1}^M = v \, \middle| M \to M \right) + P\left( v_{n+1}^M = v \, \middle| D \to M \right) \right], \\
		P_{n+1}^D\left( v \right) &= \frac{1}{2} \left[ P\left( v_{n+1}^D = v \, \middle| M \to D \, \right) + P\left( v_{n+1}^D = v \, \middle| D \to D \right) \right].
	\label{eq:Recursive_Mother_Daughter}
	\end{split} \end{equation}

	\noindent Then following Eq. \ref{eq:Post_Division_Distribution}, we can write the distributions for $v_{n+1}^M$ in terms of the growth policies $f^M(v)$ and $f^D(v)$,
	
	\begin{equation} \begin{split}
		P\left( v_{n+1}^M = v \, \middle| M \to M \right) &= \int dv' \; \left[ \iint dt^N \, dv^N \; \delta\left( v - \frac{1}{1+r} \left( f^M(v') \, e^{\lambda t^N} + v^N \right) \right) P\left(t^N, v^N \right) \right] P_n^M \left( v' \right), \\
		P\left( v_{n+1}^M = v \, \middle| D \to M \right) &= \int dv' \; \left[ \iint dt^N \, dv^N \; \delta\left( v - \frac{1}{1+r} \left( f^D(v') \, e^{\lambda t^N} + v^N \right) \right) P\left(t^N, v^N \right) \right] P_n^D \left( v' \right),
	\end{split} \end{equation}
	
	\noindent and then also for $v_{n+1}^D$,
	
	\begin{equation} \begin{split}
		P_{n+1}^D\left( v_{n+1}^D = v \, \middle| M \to D \, \right) &= \int dv' \; \left[ \iint dt^N \, dv^N \; \delta\left( v - \frac{r}{1+r} \left( f^M(v') \, e^{\lambda t^N} + v^N \right) \right) P\left(t^N, v^N \right) \right] P_n^M \left( v \right), \\
		P_{n+1}^D\left( v_{n+1}^D = v \, \middle| D \to D \, \right) &= \int dv' \; \left[ \iint dt^N \, dv^N \; \delta\left( v - \frac{r}{1+r} \left( f^D(v') \, e^{\lambda t^N} + v^N \right) \right) P\left(t^N, v^N \right) \right] P_n^D\left( v' \right).
	\end{split} \end{equation}
	
	\noindent It is again convenient to define two sub-kernels, much as we do in Eq. \ref{eq:Subkernel},
	
	\begin{equation} \begin{split}
		K^M(v,v';r) &= \iint dt^N \, dv^N \; \delta\left( v - \frac{1}{1+r} \left( f^M(v') \, e^{\lambda t^N} + v^N \right) \right) P\left( t^N, v^N \right), \\
		K^D(v,v';r) &= \iint dt^N \, dv^N \; \delta\left( v - \frac{r}{1+r} \left( f^D(v') \, e^{\lambda t^N} + v^N \right) \right) P\left( t^N, v^N \right),
	\end{split} \end{equation}
	
	\noindent with which we may rewrite Eq. \ref{eq:Recursive_Mother_Daughter} in integral form,
	
	\begin{equation} \begin{split}
		P_{n+1}^M \left( v \right) &= \frac{1}{2} \int dv' \; K^M \left( v, v'; r \right) P_n^M \left( v' \right) + \frac{1}{2} \int dv' \; K^D \left( v, v'; \frac{1}{r} \right) P_n^D\left( v' \right), \\
		P_{n+1}^D \left( v \right) &= \frac{1}{2} \int dv' \; K^M \left( v, v'; \frac{1}{r} \right) P_n^M \left( v' \right) + \frac{1}{2} \int dv' \; K^D \left( v, v'; r \right) P_n^D\left( v' \right).
	\end{split} \end{equation}
	
	Since we are looking for stable distributions, we can remove the generation dependence and rewrite the above set of equations as
	
	\begin{equation} \begin{split}
		P^M\left( v \right) &= \frac{1}{2} \int dv' \left[ K^M \left( v, v'; r \right) P^M\left( v' \right) + K^D \left( v, v'; \frac{1}{r} \right) P^D\left( v' \right) \right], \\
		P^D\left( v \right) &= \frac{1}{2} \int dv' \left[ K^M \left( v, v'; \frac{1}{r} \right) P^M\left( v' \right) + K^D \left( v, v'; r \right) P^D\left( v' \right) \right].
	\label{eq:Stable_System}
	\end{split} \end{equation}
		
	\noindent Instead of a single homogeneous Fredholm integral equation of the second kind as in Eq. \ref{eq:Eigenvalue_Equation}, we now have a \emph{system} of homogenous Fredholm integral equations of the second kind. Though these relations for mother and daughter distributions are entangled, we can decouple them. Since the asymmetry ratio $r = v^D / v^M$ is fixed, it follows that $P^M(v / r) = r \, P^D(v)$. Thus we can write the integral over $P^D(v')$ in the equation above for $P^M(v)$ as
	
	\begin{equation}
		\int dv' \; K^D \left( v, v'; \frac{1}{r} \right) P^M\left( v'\right) 
			= \int dv' \; K^D \left( v, v'; \frac{1}{r} \right) \left( \frac{1}{r} P^M\left( \frac{v'}{r} \right) \right)
			= \int dv' \; K^D \left( v, r v'; \frac{1}{r} \right) P^M\left( v' \right),
	\end{equation}
	
	\noindent which allows us to write a single Fredholm integral equation of the second kind for $P^M(v)$,
	
	\begin{equation}
		P^M\left( v \right) = \int dv' \; K^M(v,v') \, P^M \left( v' \right),
	\label{eq:Mother_Equation}
	\end{equation}
	
	\noindent where we are now able to define the mother kernel $K^M(v,v')$ as
	
	\begin{equation}
		K^M(v,v') = \frac{1}{2} \left(  K^M \left( v, v'; r \right) + K^D \left( v, rv'; \frac{1}{r} \right) \right).
	\end{equation}
	
	\noindent Then to calculate $P^D(v)$, we merely have to use the relation $P^D(v) = P^M(v/r) / r$.
	
	Now that we have a necessary condition for a stable solution (Eq. \ref{eq:Mother_Equation}), we can use it to calculate the $k$th moment,
	
	\begin{equation}
		\left\langle \left( v^M \right)^k \right\rangle = \frac{1}{2} \left( \frac{1}{1+r} \right)^k \left[ \left\langle \left( f^M(v^M) e^{\lambda t^N} + v^N \right)^k \right\rangle + \left\langle \left( f^D(v^M) e^{\lambda t^N} + v^N \right)^k \right\rangle \right] .
	\end{equation}
	
	\noindent Once again, to obtain the $k$th moment for $v^D$, we only need to note that $\left\langle \left( v^D \right)^k \right\rangle = r^k \left\langle \left( v^M \right)^k \right\rangle$. For concreteness, we again assume that both mothers and daughters follow affine linear growth policies,
	
	\begin{equation} \begin{split}
		f^M(v) &= \Delta^M + c^M \, v, \\
		f^D(v) &= \Delta^D + c^D \, v.
	\label{eq:Linear_Growth_Functions}
	\end{split} \end{equation}
	
	\noindent where the control coefficients ($c^M$ and $c^D$) and additive volumes ($\Delta^M$ and $\Delta^D$) are generally different. We can then obtain an explicit expression for the first moment:
	
	\begin{equation}
		\left\langle v^M \right\rangle = \frac{ \left( \Delta^M + \Delta^D \right) \left\langle e^{\lambda t^N} \right\rangle }{ 2(1+r) - (c^M + r c^D) \left\langle e^{\lambda t^N} \right\rangle },
	\label{eq:MD_Mean}
	\end{equation}
	
	\noindent which is positive only if
	
	\begin{equation}
		2(1+r) > (c^M + r c^D) \left\langle e^{\lambda t^N} \right\rangle.
	\end{equation}
	
	Note that, in our original model, if we were to use a purely linear policy ($\Delta = 0$), then the corresponding mean (Eq. \ref{eq:Stable_Mean}) would no longer exist (recall that the only eigenfunction in this case, $1/v$, is non-normalizable). However, by allowing mothers and daughters to follow separate policies, one of the additive volumes ($\Delta^M, \Delta^D$) can be zero without invalidating the mean (see Eq. \ref{eq:MD_Mean}).
	
	It is also useful to have an expression for the second moment:
	
	\begin{equation}
		\left\langle \left( v^M \right)^2 \right\rangle = \frac{\left(\Delta^M\right)^2 + \left(\Delta^D\right)^2 + \left( \Delta^M c^M + r \Delta^D c^D \right) \left\langle v^M \right\rangle }{2(1+r)^2 - (\left(c^M\right)^2 + r^2 \left(c^D\right)^2) \left\langle e^{2 \lambda t^N} \right\rangle} \left\langle e^{2\lambda t^N} \right\rangle,
	\end{equation}
	\end{widetext}
	
	\noindent the existence of which imposes a stronger constraint on the control coefficients $c^M$ and $c^D$,
	
	\begin{equation}
		2(1+r)^2 > \left( \left(c^M\right)^2 + r^2 \left(c^D\right)^2 \right) \left\langle e^{2\lambda t^N} \right\rangle.
	\end{equation}
	
	\begin{figure}[t!]
		\centering
			\includegraphics[width=\linewidth]{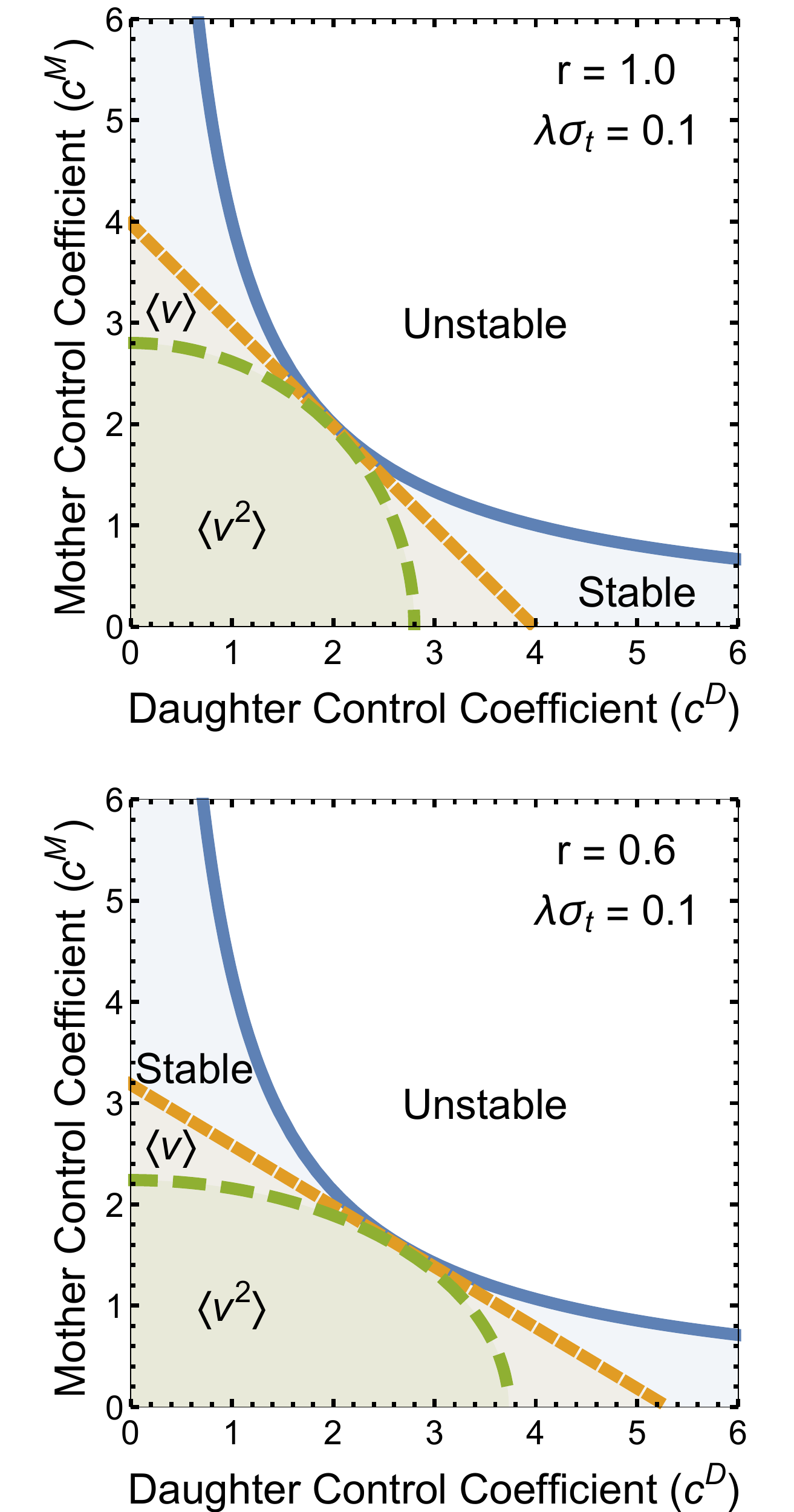}
			\caption{(Color online) Stability phase diagrams for two different asymmetry ratios. The stability phase boundary is given by Eq. \ref{eq:Phase_Boundary}. We also use our moment constraint equation (Eq. \ref{eq:MD_Moment_Constraints}) to indicate the regions in which the first moment ($\langle v \rangle$) and second moment ($\langle v^2 \rangle$) exist. Note that the $\langle v^2 \rangle$ region is a subset of the $\langle v \rangle$ region which is itself a subset of the stable region. The boundaries never intersect, though they do at some points come quite close. In both cases the noise amplitude is $\lambda \sigma_t = 0.1$. We also assume that at least one of the additive constants $\Delta^M$ and $\Delta^D$ is nonzero. \textbf{(a)} The symmetric case ($r = 1$). Note the mirror symmetry across the line $c^M = c^D$. \textbf{(b)} An asymmetric case ($r = 0.6$). The regions in which the various moments exist are skewed towards the $c^D$ axis due to the fact that daughter cells can afford to have a higher control coefficient than mother cells due to their smaller size at birth. }
	\label{fig:Growth_Factor_Constraints}
	\end{figure}
	
	We can also consider the constraints for higher moments. While the actual expression for the $k$th moment is complicated (though similar to Eq. \ref{eq:Explicit_Kth_Moment}), the constraint implied by the $k$th moment is much simpler,
	
	\begin{equation}
		2(1+r)^k > \left( \left(c^M\right)^k + r^k \left(c^D\right)^k \right) \left\langle e^{k\lambda t^N} \right\rangle.
	\label{eq:MD_Moment_Constraints}
	\end{equation}
	
	\noindent Assuming this expression is valid for non-integer $k$, we can take the limit as $k \to 0$ to determine for which values of $c^M$ and $c^D$ the system will be stable, just as we do in deriving Eq. \ref{eq:Stability_Bound}. However in this case we need to perform a few more manipulations. Assuming Gaussian multiplicative noise, we write Eq. \ref{eq:MD_Moment_Constraints} as
	
	\begin{equation}
		c^M < (1+r) \exp\left[-\frac{1}{2} k \lambda^2 \sigma_t^2 \right] \left(\frac{2}{ 1 + \left( r \frac{c^D}{c^M} \right)^k } \right)^{\frac{1}{k}}.
	\end{equation}
	
	\noindent In this form, we can take the limit as $k \to 0$ just as we do in Sec. \ref{Sec:Tail}, only this time instead of an absolute bound on the control coefficients, we obtain an equation for a phase boundary:
	
	\begin{equation}
		c^M = \frac{1+r}{\sqrt{r \frac{c^D}{c^M}}},
	\end{equation}
	
	\noindent which we can write more conveniently as
	
	\begin{equation}
		c^M = \frac{(1+r)^2}{r \, c^D}.
	\label{eq:Phase_Boundary}
	\end{equation}
	
	\noindent Note that when $c^M = c^D$, this reduces to our result from Eq. \ref{eq:Stability_Bound}.
	
	As is the case for a single-policy system, the overall stability depends only on the asymmetry ratio, not the noise amplitude. Of course stability does not require the existence of the mean or variance, and so a stable distribution is not necessarily a biologically well-behaved distribution. In Fig. \ref{fig:Growth_Factor_Constraints}, we construct a phase diagram showing the stable region for $c^M$ and $c^D$ as well as the region in which the mean exists and the region in which the variance exists for two different values of $r$. Note that these regions are concentrated along the $c^D$ axis in the asymmetric case, reflecting the fact that daughter cells can afford a larger control coefficient due to their smaller size at birth.
	
	\clearpage



\bibliographystyle{apsrev4-1}

%

\end{document}